\documentclass{birkjour} 
\usepackage{graphicx}  
\usepackage{dcolumn}   
\usepackage{amsmath}
\usepackage{amssymb}
\usepackage{amsthm}
\usepackage{color}   
\usepackage{bm}
\usepackage[backref]{hyperref}
\usepackage{enumerate}
\usepackage{transparent}
\usepackage{feynmp-auto}
\usepackage{xspace}
\usepackage{multirow,bigdelim}

\def\fmfsettings{
	\fmfset{thin}{0.7pt}
	\fmfset{arrow_len}{6pt}
}

\def\cD{\mathcal{D}}
\def\cG{\mathcal{G}}
\def\cT{\mathcal{T}}

\def\cM{\mathcal{M}}

\def\IR{\mathbb{R}}
\def\IC{\mathbb{C}}
\def\IZ{\mathbb{Z}}
\def\nn{\nonumber}
\def\Ys{\ensuremath{\mathcal{Y}_S}}
\def\ap{{\alpha'}}
\def\apt{\ensuremath{\ap \to 0}}
\def\trop{\mathrm{trop}}
\def\tach{\mathrm{tach}}

\def\Re{\mathrm{Re\,}}
\def\Im{\mathrm{Im\,}}

\def\cH{\mathcal H}

\def\M{\cM_{g,n}}
\def\Mbar{\overline{\cM}_{g,n}}
\def\Mgbar{\overline{\cM}}
\def\Mt{\cM_{g,n}^\trop}

\newcommand{\Mtgn}[2]{\cM_{#1,#2}^\trop}

\newcommand{\Mgn}[2]{\cM_{#1,#2}}

\newcommand{\sM}[2]{{\mathfrak M}_{#1,#2}}

\def\ab{\left[ \begin{smallmatrix} \bm{\beta}\\
      \bm{\alpha} \end{smallmatrix} \right]}
\newcommand{\charac}[2]{\left[ \begin{smallmatrix} #1\\
      #2 \end{smallmatrix} \right]}

\def\cDempty{\mathcal D_0}

\def\dmubos{{\mathrm d} \mu_\mathrm{bos} \,}
\def\dmutrop{\mathrm d\mu_\trop\,}

\def\d{\mathrm d}
\newcommand{\Ast}[2]{\mathsf A^{(#1,#2)}_\ap}
\newcommand{\A}[2]{\mathsf A^{(#1,#2)}}

 \newtheorem{thm}{Theorem}
 \newtheorem{lem}{Lemma}
 \newtheorem{prop}[thm]{Proposition}
 \newtheorem{conj}[thm]{Conjecture}

\def\BKcite{\cite{Bern:1990cu,Bern:1990ux,Bern:1991aq,Bern:1993wt}\xspace}

\numberwithin{equation}{section}

\begin{document}


\title{Tropical Amplitudes}
\author{Piotr Tourkine}
\address{Department of Applied Mathematics and Theoretical Physics,\\
Wilberforce Road, Cambridge CB3 0WA,
United Kingdom,}
\email{pt373@cam.ac.uk}

\begin{abstract}
  In this work, we argue that the $\alpha'\to 0$ limit of closed string
  theory scattering amplitudes is a tropical limit.  
  The motivation is to develop a technology to systematize the
  extraction of Feynman graphs from string theory amplitudes at higher
  genus.
  An important technical input from tropical geometry is the use of
  tropical theta functions with characteristics to rigorously derive
  the worldline limit of the worldsheet propagator.
  This enables us to perform a non-trivial computation at two loops:
  we derive the tropical form of the integrand of the genus-two
  four-graviton type II string amplitude, which matches the direct
  field theory computations.
  At the mathematical level, this limit is an implementation of the
  correspondence between the moduli space of Riemann surfaces and the
  tropical moduli space.
\end{abstract}


\maketitle
\tableofcontents

\section{Introduction}
\label{sec:intro}
It is well accepted that the field theory limit\footnote{Throughout
  the text, we call indistinctly, ``point-like'', ``field theory'',
  ``infinite tension'' ``tropical'' or ``$\apt$'' this limit. We
  recall that the Regge slope $\ap$ of the string is a positive
  quantity of mass dimension $-2$ related to the string length
  $\ell_s$ by $\ap = \ell_s^2$.} of string theory scattering
amplitudes reproduces the usual perturbative expansion of quantum
field theory. However a constructive general proof of that statement has
not been given yet. Besides the intrinsic interest of such a proof,
this problem is important for several reasons.

Firstly, string inspired methods have already proved their efficiency at
one loop to compute scattering amplitudes in field theory
\cite{Green:1982sw,Metsaev:1987ju,Kaplunovsky:1987rp,Minahan:1987ha,
Dixon:1990pc,Bern:1987tw,Bern:1990cu,Bern:1990ux,Bern:1991aq,Bern:1993wt,
Dunbar:1994bn,DiVecchia:1996kf,DiVecchia:1996uq,Green:1999pv,Schubert:2001he}
and to obtain more general results about amplitudes
\cite{BjerrumBohr:2008ji,BjerrumBohr:2008vc,Badger:2008rn,Schubert:2010tx,
Schubert:2011fz,Dunne:2005sx,Dunne:2006st}. 
Secondly, it is important to better understand the mechanisms by which
string theory renormalizes supergravity theories. In particular, the
question of the ultraviolet (UV) divergences of maximal supergravity
continues to draw much attention
\cite{Bern:2007hh,Bern:2008pv,Bern:2009kd,
  Bossard:2009sy,Vanhove:2010nf,Bjornsson:2010wm,Bossard:2010bd,
  Bossard:2011tq,Kallosh:2011dp,Berkovits:2009aw,Green:2010sp} and
string theory provides a well-suited framework to analyse this issue
\cite{Berkovits:2009aw,Green:2010sp,Green:2010kv,Green:2010wi,Green:2011vz}.

In this paper, we revisit the $\apt$ limit of string
theory~\cite{Scherk:1971xy} in the context of tropical geometry, a
link previously unnoticed. Since tropical geometry describes -- in
particular -- how Riemann surfaces degenerate to certain graphs called
tropical graphs, it provides a framework for studying this
limit. Tropical graphs are then seen as particles' worldlines.

Only at one-loop the Bern-Kosower rules \BKcite~give a full-fledged
method to obtain field theory amplitudes from string theory. At higher
loops such techniques are not available and this work is a step in
this direction.\footnote{An alternative approach exists in the
  literature to study the $\apt$ limit of string amplitudes, based on
  the Schottky parametrization, see the recent works
  \cite{Magnea:2013lna,Magnea:2015fsa}.}

The aim of this work is therefore computational: it is to develop
methods based on tropical geometry to extract the field theory limit
of higher genus closed string theory amplitudes.

\medskip

The ``tropicalization'' of a complex variety is a particular
degeneration by which the variety sees its dimension halved. Consider
for instance the annulus $\Sigma=\{z,1<|z|<\rho\}$.
The \textit{tropical} variety is obtained by a taking the ``modulus''
of the coordinate in $\Sigma$; paraphrasing~\cite{Itenberg:2011}, the
tropical limit corresponds to ``\textit{forgetting the phases in
  complex numbers}''.  The meaning of the modulus of $z$ is easier
seen by mapping the annulus to the cylinder via $z\to \exp{i w}$ with
$w=\sigma_1+i \sigma_2$: $|z|$ is a longitudinal coordinate along the
cylinder and the tropical variety is just a segment in this case.

We will make this more precise for generic Riemann surfaces in
sec.~\ref{sec:class-geom-trop}. It should however already be clear
that this process is similar to the point-like limit of string
theory. Seeing the cylinder as the worldsheet of a closed string
propagating through spacetime, the phase-dependence of the amplitude
enforces the ``level-matching'' condition. Level-matching is a
physical constraint that forces the string to be balanced and have as
many left-moving as right-moving excitations. But, in the $\apt$
limit, one could think that the massive excitations, that have masses
of order $1/\alpha'$, should decouple and make the level matching
condition trivial. There is however a caveat. When the field theory
amplitudes have ultraviolet (UV) divergences, the massive modes do not
decouple but instead act as UV regulators. These give rise to
counter-terms in the amplitudes. We shall see that these counter-terms
have a natural description in tropical geometry: they correspond to
certain weighted vertices.

\medskip 

This text begins in sec.~\ref{sec:tropgeom} with an introduction to
tropical geometry. In sec.~\ref{sec:trop-theta-char}, we prove an
important lemma, on tropical theta functions with characteristics,
lem.~\ref{lem:dist}. Later we make use of it to show that the $\apt$
limit of the string theory propagator on higher genus surfaces reduces
to the worldline propagator.
This tropical limit of the string propagator is one of the main
contributions of this work. This step is required to extract in full
rigor the form of the field theory amplitudes arising in the $\apt$
limit of string theory.  This discussion is extended in
sec.~\ref{sec:class-geom-trop} to the connection between tropical and
classical geometry.

In sec.~\ref{sec:ftl}, we formulate the field-theory limit of closed
string theory amplitudes in the context of tropical geometry.
We explain how, as \apt, a genus $g$, $n$-point string theory
amplitude $\A g n _\ap$ reduces to an integral over the moduli space
of tropical graphs ~\cite{Brannetti:2009,Caporaso:2010}, $\Mt$
\begin{equation}
  \lim_{\apt} \A g n _\ap = \int_{\Mt} \d\mu^\trop F_{g,n}\,,
 \label{e:tropamps}
\end{equation}
The right-hand side of this equation is the renormalized field theory
amplitude written in its ``tropical representation'', or in short a
``tropical amplitude''.  The integration measure $\d \mu^\trop$ is
defined in terms of the Schwinger proper times of the graph -- the
lengths of the inner edges. The integrand $F_{g,n}$ contains the
theory-dependence of the amplitude and encompasses both the numerators
and denominators of the Feynman graphs (see eq.~\eqref{prop:sym}
below). This type of formulas are the origin of Feynman's construction
of quantum field theory~\cite{Feynman:1950ir}. The novelty of our
approach lies in the use of tropical geometry to extract the limit, which
allows to recycle some of the string theory efficiency and compactness
in field theory.

We come to practical applications in sec.~\ref{sec:troppf}. We start
with a review of tree-level and one-loop methods. Then we compute the
tropical limit of the two-loop four-graviton type~II string amplitude
of D'Hoker and Phong
\cite{D'Hoker:2001nj,D'Hoker:2001zp,D'Hoker:2001it,D'Hoker:2002gw,
  D'Hoker:2005jb,D'Hoker:2005jc,D'Hoker:2007ui} and find agreement
with the supergravity result of \cite{Bern:1998ug,Green:2008bf}; that
is another main contribution of this paper.

Besides the study of the $\apt$ limit of string amplitudes, our
approach sheds a new light on the geometry of field theory amplitudes:
they are integrals over the tropical moduli space. The components of
the Feynman integrands also acquire a geometrical origin: the first
Symanzik polynomial is seen to be the determinant of the period matrix
of the tropical graph, while the second is written in terms of Green's
functions on the graph.  Similar observations were made
in~\cite{Dai:2006vj,Green:2008bf}.

We close this introduction with a comment. String field theory
constructions, Zwiebach's bosonic string field theory in
particular~\cite{Zwiebach:1992ie}, give formal representations of
string field theory amplitudes in terms of certain Feynman
graphs. Although massless fields (field theory fields) contributions
are accounted for in these graphs, these constructions are not
designed for practical implementation of the field theory limit.
Their goal is rather a non-perturbative formulation of string field
theory.  In principle one could take formally the $\apt$ limit of a
string field theory amplitude. This would lead us to a set of Feynman
rules and a prescription to build field theory amplitudes: the exact
same one as if we had started with a field theory Lagrangian.

What we want to do here is the opposite. We want to be able to take a
string theory amplitude, expressed in its compact form as a single
moduli space integral, and extract field theory graphs out of it, in
the spirit of the Bern-Kosower rules.

\textbf{Note added.}\textit{ In the second version of this paper the
  author added a comment on the three-loop amplitude of
  \cite{Gomez:2013sla} at the end of sec.~\ref{sec:troppf}.}

\section{Tropical geometry}
\label{sec:tropgeom}

Tropical geometry is a recent and active field in
mathematics.\footnote{For introductory works, the reader is referred
  to \cite{Itenberg:2011,Mikhalkin:2004,Speyer:2004,Mikhalkin:2006,
    Mikhalkin:2007,Caporaso:2011a,Mikhalkin:2006a}, and to
  \cite{Brannetti:2009} for a more exhaustive bibliography.}  The
basic objects, tropical varieties, can be either abstract
\cite{Baker:2007} or defined as algebraic curves over certain spaces
\cite{Mikhalkin:2006}. Tropical varieties also arise as the result of
a degeneration of the complex structure of complex varieties called
\textit{tropicalization} \cite{Abramovich:2012,Abramovich:2013}.
The use of tropical geometry in physics is not new: even before the
coinage of the word ``tropical'', the authors of \cite{Aharony:1997bh}
studied a class of embedded tropical varieties called webs, arising
from the degeneration of brane models.  Also, Kontsevich and Soibelman
introduced tropical geometry in the context of mirror symmetry
\cite{Kontsevich:2004}, which became an active area of investigation
(see the survey \cite{mark2011tropical}).

\subsection{Tropical graphs}
\label{sec:abstropcurv}
\begin{figure}[t]
	\centering
            \parbox{30pt}{ 
		\begin{fmffile}{n3}
\fmfsettings
		\fmfframe(10,0)(0,0)
		{ 
		\begin{fmfgraph*}(30,30)
			\fmflabel{$1$}{g1}
			\fmflabel{$2$}{g2}
			\fmflabel{$3$}{g3}
			\fmfleft{g1,g2}
			\fmfright{g3}
			\fmf{plain}{g1,v}
			\fmf{plain}{g2,v}
			\fmf{plain,tension=1.3}{g3,v}
		\end{fmfgraph*} }
		\end{fmffile} }
	\hspace{48pt}
		\parbox{30pt}{ \begin{fmffile}{n1g1}
\fmfsettings
		\fmfframe(0,0)(0,0){ \begin{fmfgraph*}(50,50)
		\fmfleft{g1}
		\fmfright{g4}
		\fmflabel{$1$}{g1}
		\fmf{plain,tension=1.5}{g1,v1}
		\fmf{phantom,tension=1.5}{g4,v4}
		\fmf{plain,left,tension=0.5}{v1,v4}
		\fmf{plain,right,tension=0.5}{v1,v4}
      \end{fmfgraph*} }
      \end{fmffile} }
      \hspace{20pt}
	\parbox{30pt}{ 
		\begin{fmffile}{n0g2}
\fmfsettings
		\fmfframe(0,0)(0,0){ 
			\begin{fmfgraph*}(40,32)
			\fmfleft{g1,g2}
			\fmfright{g4,g3}
			\fmf{phantom}{g1,v1}
			\fmf{phantom}{g2,v2}
			\fmf{phantom}{g3,v3}
			\fmf{phantom}{g4,v4}
			\fmf{phantom,tension=0.5}{v1,v2}
			\fmf{phantom,tension=0.5}{v2,v3}
			\fmf{phantom,tension=0.5}{v3,v4}
			\fmf{phantom,tension=0.5}{v4,v1}
			\fmf{phantom,left,tension=0.,tag=1}{v1,v3}
			\fmf{phantom,left,tension=0.,tag=2}{v3,v1}
			\fmf{phantom,left,tension=0.,tag=1}{v2,v4}
			\fmf{phantom,left,tension=0.,tag=2}{v4,v2}
			\fmfposition
			\fmfipath{p[]}
			\fmfiset{p1}{vpath1(__v1,__v3)}
			\fmfiset{p2}{vpath2(__v3,__v1)}
			\fmfi{plain}{subpath (0,length(p1)/2) of p1}
			\fmfi{plain}{subpath (length(p1)/2,length(p1)) of p1}
			\fmfi{plain}{subpath (0,length(p2)) of p2}
			\fmfi{plain}{point 3length(p1)/4 of p1 -- point 3length(p2)/4
of p2}
		\end{fmfgraph*} }
		\end{fmffile} }
	\hspace{12pt}
			\parbox{40pt}
{ \begin{fmffile}{n0g1}
\fmfsettings
		\fmfframe(0,0)(0,0){ \begin{fmfgraph*}(50,50)
		\fmfleft{g1}
		\fmfright{g4}
		\fmf{phantom,tension=1.5}{g1,v1}
		\fmf{phantom,tension=1.5}{g4,v4}
		\fmfv{decor.shape=circle,decor.filled=full,
		decor.size=1thick,label.dist=3pt}{v4}
		\fmflabel{$w, w>0$}{v4}
		\fmf{plain,left,tension=0.5}{v1,v4}
		\fmf{plain,right,tension=0.5}{v1,v4}
      \end{fmfgraph*} }
      \end{fmffile} }\\
      \vspace{-8pt}
\caption{Examples of tropical graphs (left to right): a $3$-point
tropical tree, a once-punctured graph of genus one, a 2-loop tropical
graph, a graph of genus $1+w$.} 
\label{fig:extropcurv}
\end{figure}
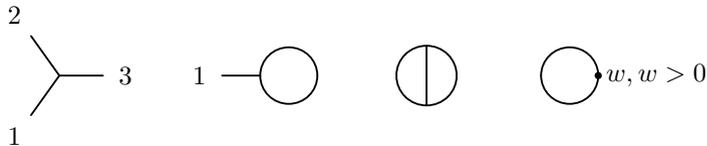

An abstract {tropical graph} is a connected graph with labeled legs (external
edges), whose inner edges have a length and whose vertices are
weighted.  The external legs are called {punctures} or marked points
and they have infinite length.
A tropical graph $\Gamma$ is then a triple $\Gamma=(G,w,\ell)$ where;
$G$ is a connected graph called the {combinatorial type} of $\Gamma$,
$\ell$ and $w$ are length and weight functions on the edges and on the
vertices
\begin{equation}
	\begin{aligned}
 \ell&:E(G)\cup L(G)\rightarrow \IR_+ \cup \{\infty\}\,,\\
 w&:V(G)\rightarrow \IZ_+\,.
	\end{aligned}
\end{equation}
The quantities $E(G)$, $L(G)$ and $V(G)$ are respectively the sets of
inner edges, legs and vertices of the graph. The {total weight} $|w|$
of a tropical graph $\Gamma$ is the sum of all the weights of the
vertices $|w|=\sum_{V(G)} w(V)$. Its {genus} $g(\Gamma)$ is the number
of loops $g(G)$ of $G$ plus the total weight
\begin{equation}
 g(\Gamma)=g(G)+|w|\,.
 \label{e:genustrop}
\end{equation}
A \textit{pure} tropical graph is by definition a tropical graph that
only has vertices of weight zero, therefore its genus of is given by
the number of loops in the usual sense. In fig.~\ref{fig:extropcurv}
we give a few examples of tropical graphs.

As for classical complex curves, a stability condition must be added
to the previous definitions; we consider only genus-$g$ tropical
graphs with $n$ punctures for which\footnote{Strictly speaking, the
  local valency condition should be viewed as considering
  \textit{classes} of abstract tropical graphs under the equivalence
  relation that contracts edges connected to 1-valent vertices of
  weight~$0$, and removes weight~$0$ bivalent vertices. Physically, on
  the worldline, this equivalence relation is perfectly sensible,
  since no interpretation of these 1- or 2- valent vertices of weight
  zero seem natural in the absence of external classical sources.}
\begin{equation}
 2g-2+n \geq 1\,.
\label{e:stabcond}
\end{equation}
This implies that every vertex of weight zero must have valency at least
three and vertices of weight one should have at least one leg.

\begin{figure}[b]
\begin{center}
	\parbox{60pt}{ 
		\begin{fmffile}{spec1}
\fmfsettings
		\fmfframe(0,0)(0,0){ \begin{fmfgraph*}(50,50)
		\fmfleft{g1}
		\fmfright{g2prime,g2,g3,g4,g5}
		\fmf{phantom,tension=1.5}{g1,v1}
		\fmf{plain,tension=1}{g4,v4}
		\fmf{plain,tension=1}{g2,v4}
		\fmf{plain,tension=1}{g3,v4}
		\fmfv{decor.shape=circle,decor.filled=full,
		decor.size=2thick
		,label.dist=-12pt
		}{v4}
		\fmflabel{$w$}{v4}
		\fmf{plain,left,tension=0.5,label=$t$}{v1,v4}
		\fmf{plain,right,tension=0.5}{v1,v4}
      \end{fmfgraph*} }
      \end{fmffile} }
      $\longrightarrow$
	\parbox{60pt}{ 
		\begin{fmffile}{spec2}
\fmfsettings
		\fmfframe(0,0)(0,0){ \begin{fmfgraph*}(50,50)
		\fmfleft{g1}
		\fmfright{g2prime,g2,g3,g4,g5}
		\fmf{phantom,tension=1.5}{g1,v4}
		\fmf{plain,tension=1}{g4,v4}
		\fmf{plain,tension=1}{g2,v4}
		\fmf{plain,tension=1}{g3,v4}
		\fmfv{decor.shape=circle,decor.filled=full,
		decor.size=2thick
		,label.dist=-32pt
		}{v4}
		\fmflabel{$w+1$}{v4}
      \end{fmfgraph*} }
      \end{fmffile} }
      \hspace{24pt}
      	\parbox{60pt}{ 
		\begin{fmffile}{spec3}
\fmfsettings
		\fmfframe(0,0)(0,0){ \begin{fmfgraph*}(50,50)
		\fmfleft{g0,g1,g2,g3,g3prime}
		\fmfright{g4prime,g4,g5,g6,g7}
		\fmf{plain}{g1,v1}
		\fmf{plain}{g2,v1}
		\fmf{plain}{g3,v1}
		\fmf{plain}{g4,v4}
		\fmf{plain}{g5,v4}
		\fmf{plain}{g6,v4}
		\fmf{plain,tension=1.2,label=$t$}{v1,v4}
		\fmfv{decor.shape=circle,decor.filled=full,
		decor.size=2thick
		,label.angle=110
		}{v4}
		\fmfv{decor.shape=circle,decor.filled=full,
		decor.size=2thick
		,label.angle=70
		}{v1}
		\fmflabel{$w_1$}{v1}
		\fmflabel{$w_2$}{v4}
      \end{fmfgraph*} }
      \end{fmffile} }
	$\longrightarrow$\hspace{8pt}
	\parbox{60pt}{ 
		\begin{fmffile}{spec4}
\fmfsettings
		\fmfframe(0,0)(0,0){ \begin{fmfgraph*}(50,50)
		\fmfleft{g0,g1,g2,g3,g3prime}
		\fmfright{g4prime,g4,g5,g6,g7}
		\fmf{plain}{g1,v1}
		\fmf{plain}{g2,v1}
		\fmf{plain}{g3,v1}
		\fmf{plain}{g6,v1}
		\fmf{plain}{g5,v1}
		\fmf{plain}{g4,v1}
		\fmfv{decor.shape=circle,decor.filled=full,
		decor.size=2thick
		,label.angle=-90
		,label.dist=10pt
		}{v1}
		\fmflabel{{\small$w_1+w_2$}}{v1}
      \end{fmfgraph*} }
      \end{fmffile} }
\caption{Specialization rules as $t\to0$.} 
\label{fig:specializations}
\end{center}
\end{figure}
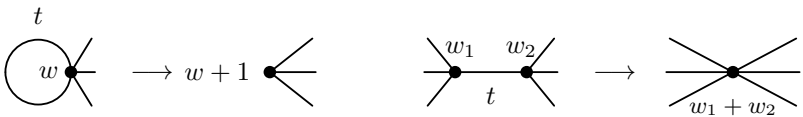

A specialization map acts on these graphs by contracting edges and
adding the weights of the vertices that are brought together, as
pictured in fig.~\ref{fig:specializations}. This gives another
interpretation of the weights; they correspond to degenerated loops,
and it is easily checked that the genus of a graph \eqref{e:genustrop}
and the stability criterion \eqref{e:stabcond} are stable under
specialization.

Finally, a graph that can be disconnected in two components by
removing a single edge is called {one-particle-irreducible} (1PI),
otherwise it is called {one-particle-reducible} (1PR).

Physically, tropical graphs will be interpreted as the worldlines
swept by propagating particles, just like Riemann surfaces are strings
worldsheets. The lengths of the edges are Schwinger proper times, and
a nonzero weight on a vertex indicates the possible insertion of a
counter-term to a divergence in the graph. Since loops with very short
proper times correspond to the UV region, it is intuitively clear that
this should be the case. In particular, at genus $g$, the tropical
graph corresponding to single vertex of weight $g$ will be supporting
counter-terms to the primary divergence of the amplitude.

\subsection{Homology, forms, Jacobian and divisors}
\label{sec:jactrop}

In this paragraph, following \cite{Mikhalkin:2006}, we introduce the
tropical analogues of some common objects of classical geometry;
abelian forms, period matrices and Jacobian varieties. Some care is
required because graphs of identical genus may not have the same
number of inner edges. We first avoid this subtlety and start with
pure graphs.

\begin{figure}[t]
\centering
  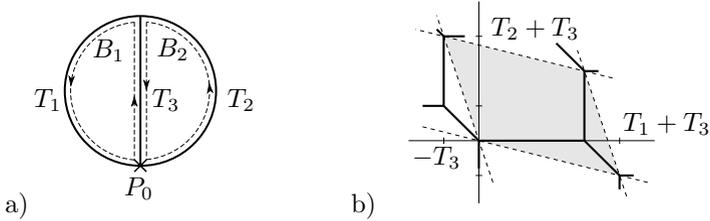
\caption{a) Genus two graph with edges lengths
  $T_1,T_2,T_3$. b)~Image of $\Gamma$ (thick line) by the tropical Abel-Jacobi map
in the Jacobian $J(\Gamma)=\IR^2/K^{(2)}\IZ^2$.}
\label{fig:jacgen2}
\end{figure}

Let $\Gamma$ be a pure tropical graph of genus $g$
and $(B_1,\ldots,B_g)$ be a canonical homology basis of $\Gamma$, as
in fig.~\ref{fig:jacgen2}~a.
The vector space of the $g$ independent tropical one-forms
$\omega_I^\trop$ can be canonically defined by;
\begin{equation}
 \omega_I^\trop = 
 \begin{cases}
   1\ \mathrm{on\ }B_I\,,\\
   0\ \mathrm{otherwise}\,.
 \end{cases}
\label{e:trophd}
\end{equation}
These forms are \textit{constant} on the edges of the graph.
The period matrix $\bm K$ is a $g\times g$ positive definite
real-valued matrix, defined by
\begin{equation}\label{e:tropK}
 \oint_{B_I} \omega^\trop_J = K_{IJ}\,.
\end{equation}
The Jacobian of $\Gamma$ is a {real torus} defined by
\begin{equation}
\label{e:tropjac}
J(\Gamma)=\IR^{g}/\bm{K}\IZ^{g}\,.
\end{equation}
The {tropical version of the Abel-Jacobi map}
$\mu^\trop$~\cite{Mikhalkin:2006,Baker:2007} is then defined by
integration along a path $\gamma$ between $P_0$ and $P_1$ on the
graph as a map to $J(\Sigma)$;
\begin{equation}
 \mu^\trop(P_0,P_1)= \int_\gamma {(\omega_1^\trop,\ldots,
\omega_g^\trop)} \mod \bm K \IZ^g\,.
\end{equation}
Changing $\gamma$ by elements of the homology basis results in adding to the
integral in the right-hand side some elements of the lattice $K \IZ^g$. Thus
$\mu^\trop$ is well defined as a map to the Jacobian torus. 
Here are two examples taken from~\cite{Mikhalkin:2006}.\\
\textit{Example 1.} Let $\Gamma$ be the genus two tropical graph depicted in
figure~\ref{fig:jacgen2} a) with canonical homology basis as in figure
\ref{fig:jacgen2}. Its period matrix is
\begin{equation}
 \bm K^{(2)}=
\begin{pmatrix}
  T_1+T_3 & -T_3 \\
  -T_3 & T_2+T_3
 \end{pmatrix}\,.
\label{e:permatrixex1}
\end{equation}
Choosing $P_0$ as depicted, one can draw the image of $\Gamma$ by the tropical
Abel-Jacobi map in $J(\Gamma)$, as shown in the figure \ref{fig:jacgen2}
b).

\begin{figure}[b]
\centering
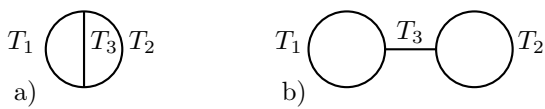
\caption{Genus-two graphs described in the examples.}
\label{fig:gen2graphs}
\end{figure}

\noindent\textit{Example 2.} Fig.~\ref{fig:gen2graphs} depicts two
inequivalent pure tropical graphs of genus two. The period matrix
$K^{(2)}$ of the 1PI graph a) is given in \eqref{e:permatrixex1} while
that of the 1PR graph b) is given by $\mathrm{Diag}(T_1,T_2)$. This
illustrates the fact that the period matrix is independent of the
lengths of the separating edges.

The generalization of this discussion to the case of tropical graphs
with weighted vertices depends on the approach one wants to use. A
simplistic approach consists in using a homology basis of size $g(G)$
instead of $g(\Gamma)$, thereby ignoring the weights on the vertices;
in this case the definitions given before apply
straightforwardly. However, in doing so, the dimension of the Jacobian
drops under specialization. A more complete treatment of this question
is provided in ref.~\cite{Brannetti:2009}.

\subsection{Divisors and theta characteristics}
Now we introduce the notion of divisors and rational functions in
order to define tropical theta characteristics. 
\subsubsection{Divisors on graphs}
\label{sec:divisors-graphs}

A divisor $D$ on a tropical graph is a formal sum of points, weighted
by integer multiplicities;
\begin{equation}
  \label{eq:divisor}
  D=\sum_{i=1}^{n}a_{i}P_{i}\,,\quad a_{i}\in\mathbb{Z}\,.
\end{equation}
The degree of a divisor is given by the sum of its weights; in the
previous example it is $a_{1}+\cdots + a_{n}$.

\begin{figure}[t]
  \centering
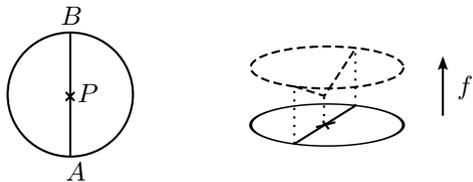
\caption{Example of rational function $f$ on a two-loop
  graph.}
\label{fig:rat-fun}
\end{figure}

A rational function on a tropical graph is a continuous,
piecewise-linear function with integer slopes (see
fig.~\ref{fig:rat-fun}). The order of a rational function at a divisor
$P$ is defined by the sum of the outgoing slopes at $P$. A rational
function is said to have a pole of order $n$ at $P$ if its order is
$-n<0$. It is said to have a zero of order $n$ if its order is
$n>0$. For $n=0$, the function is simply regular at
$P$.\footnote{Strictly speaking, another property should be added to
  the definition of a rational function: it must have finitely many
  poles and zeros. Thus, a rational function has finitely many linear
  pieces.}

The divisor $\mathrm{div}(f)$ of a rational function $f$ is defined to
be the sum of the divisors $P$ of the graph, weighted by the order of
$f$ at $P$. In the example of fig.~\ref{fig:rat-fun}, if the slopes of
the $f$ on the central edge are $\pm1$, then we find
$\mathrm{div}(f)=2P -A-B$.

\begin{figure}[b]
  \centering
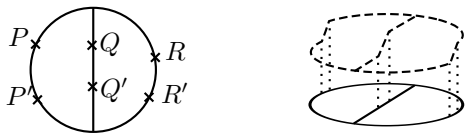
\caption{Example of linear equivalence; $P+Q+R\sim P'+Q'+R'$}
\label{fig:lin-eq}
\end{figure}

Two divisors $D$ and $D'$ are said to be linearly equivalent,
$D\sim D'$, if and only if there exists a rational function $f$ whose
divisor is $D-D'$, as in fig.~\ref{fig:lin-eq}.
Finally, a canonical divisor on a graph is a linear equivalence class
of divisors $D$ of which a representative $K_{\Gamma}$ is defined by
\begin{equation}
  \label{eq:candiv}
  K_{\Gamma}=\sum_{P\in \Gamma}(\mathrm{valence}(P)-2) P\,.
\end{equation}
For instance, if $\Gamma$ is a trivalent graph, a representative
canonical divisor is the sum of the points at the vertices; on the
example of fig.~\ref{fig:rat-fun}, $K=A+B$.

\subsubsection{Tropical theta characteristics}
\label{sec:trop-theta-char}

To define tropical theta characteristics, originally introduced in
\cite{Mikhalkin:2006,Zharkov:2006}, we follow
\cite{2014arXiv1404.7568B}. A theta characteristics on a graph
$\Gamma$ is a class of divisors $D$ such that $2D$ is linearly
equivalent to $K_{\Gamma}$;
\begin{equation}
  \label{eq:trop-theta-char}
  2D\sim K_{\Gamma}
\end{equation}

This definition is equivalent to the following. To define a theta
characteristics on a graph $\Gamma$, first define a $\mathbb{Z}_{2}$
flow on the graph, i.e. a cycle $C$ on $\Gamma$ (possibly
disconnected) such that at each vertex the number of edges belonging
to the cycle is 0 modulo~2. Then put arrows on the complement of $C$
in $G$ that go in the direction opposite to $\Gamma$.
Where the arrows meet, insert a divisor weighted by the numbers of
edges meeting there, minus 1. Then, this divisor is a theta
characteristics in the sense of eq.~\eqref{eq:trop-theta-char}, as
shown in refs.~\cite[Lemma 6]{Zharkov:2006} or \cite[Lemma 3.4]{2014arXiv1404.7568B}.
Different choices of flows produce non-equivalent tropical
theta characteristics. In total, there are $2^g$ tropical theta
characteristics~\cite{Zharkov:2006}.

\begin{figure}[t]
  \centering
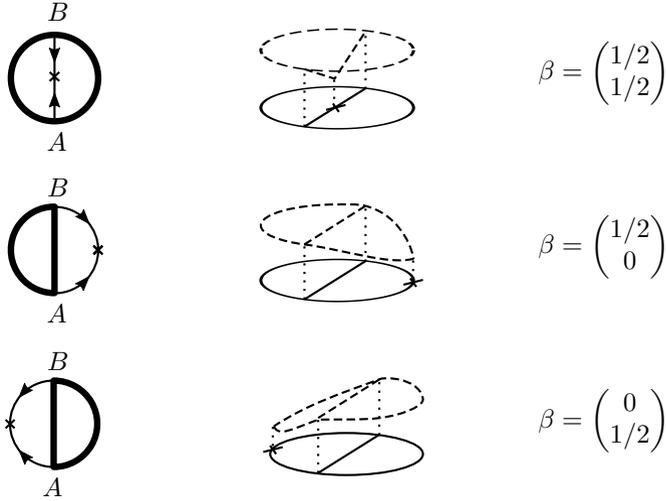
\caption{The three tropical theta characteristics at genus two.}
\label{fig:theta-char-2-loop}
\end{figure}

While the relation between tropical and classical theta
characteristics does not appear to have been discussed in the
literature, we will here conjecture how to associate a $g$-dimensional
vector to a tropical theta characteristics. 

Take the flow $C$ defined above, it is uniquely decomposed in the
homology as
\begin{equation}
  \label{eq:flow-decomp}
  C=\cup_{i\in \mathcal{I}}B_{i}\,,
\end{equation}
for some unique set $\mathcal{I}$.
It is then conjectured here that the theta characteristics associated
to this cycle is the vector $\bm{\beta}$ of
$\tfrac12 (\mathbb{Z}/2\mathbb{Z})^{g}$ with entries $\beta_{i}$,
$i=1,\ldots, g$ such that
\begin{equation}
  \label{eq:trop-theta-beta}
  {\beta}_{i}=  
       \begin{cases}
1/2\ \mathrm{if}\ i\in \mathcal{I}\,,\\
0\ \mathrm{otherwise}\,.
       \end{cases}
     \end{equation}
An example of this construction is provided in fig.~\ref{fig:theta-char-2-loop}.

We now have the following lemma.

\begin{lem}
  Let $P$ and $Q$ be two points on a tropical graph $\Gamma$, let
  $\gamma$ be a path joining them and $\mathrm{dist}_{\gamma}(P,Q)$ be
  the distance between $P$ and $Q$ along $\gamma$. Then, there always
  exist a tropical theta characteristics
  $\bm \beta\in \tfrac12 (\mathbb{Z}/2\mathbb{Z})^{g}$ such that
\begin{equation}
  \label{eq:char-lem}
  \bm \beta \cdot \int_{\gamma} (\omega^{\trop}_{1}\ldots
  \omega^{\trop}_{g}) = \frac12\rm{dist}_{\gamma}(P,Q)\,.
\end{equation}
\label{lem:dist}
\end{lem}

\paragraph{Proof.} First, given two points $P$ and $Q$ joined
by a path $\gamma$, there always exist at least one $\mathbb{Z}_{2}$
flow $C$ containing $\gamma$. This cycle is decomposed
uniquely as a particular union of homology cycles; this defines a
corresponding set $\mathcal{I}_{C}$, as in \eqref{eq:flow-decomp}.

Let $\bm \beta^{(C)}$ be the tropical theta characteristics
associated to $C$ as in eq.~\eqref{eq:trop-theta-beta}.
By definition, its only nonzero entries $\beta^{(C)}_{J}\neq0$ are these for which
$J\in \mathcal{I}_{C}$.
The entries of the vector
$\int_{\gamma} (\omega^{\trop}_{1},\ldots, \omega^{\trop}_{g})$, into
which $\bm \beta^{(C)}$ is dotted, result from the integration of the
tropical one-forms along $\gamma$.
By definition again, the individual one-forms $\omega_{J}^{\trop}$
integrated along $\gamma$ give exactly the length of the portion of
the cycle $B_{J}$ that belongs to $\gamma$, which we can call
$\gamma_{J}$.
Note that if $\gamma\cap B_{J}=\emptyset$, then
$\gamma_{J}=0$. In general, several cycles share an edge
$\gamma\cap B_{J_{1}}=\ldots=\gamma\cap B_{J_{k}}$ and this implies
that the vector
$\int_{\gamma} (\omega^{\trop}_{1},\ldots, \omega^{\trop}_{g})$ has
entries that can be equal.

The scalar product with $\bm \beta^{(C)}$ precisely has the effect to
avoid to double count these components.  
Indeed, amongst all these cycles $B_{J_{1}},\ldots B_{J_{k}}$ which
would produce identical terms, the unique
decomposition \eqref{eq:flow-decomp} picks only the one that belongs
to $\mathcal{I}_{C}$. Therefore, the left-hand side of
\eqref{eq:char-lem} is rewritten as the following sum
\begin{equation}
  \label{eq:sumgammaJ}
  \begin{aligned}
    \bm\beta^{(C)} \cdot \int_{\gamma} (\omega^{\trop}_{1}\ldots
    \omega^{\trop}_{g})
    &= \sum_{J=1}^{g}\beta^{(C)}_{J} \gamma_{J}\\
    &=\frac 12 \sum_{J\in \mathcal{I}_{C}}\gamma_{J}
  \end{aligned}
\end{equation}
where the right-hand side of the second line is one-half of the length
of the path $\gamma$, as claimed.
\hfill $\square$

\begin{figure}[t]
  \begin{center}
    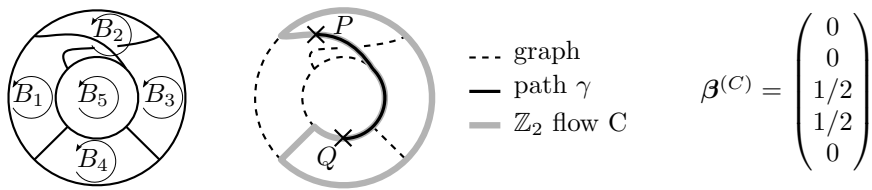
    \caption{Five-loop tropical characteristics and illustration of
      the lemma.}
    \label{fig:fiveloop}
  \end{center}
\end{figure}

Fig.~\ref{fig:fiveloop} shows an illustration of this proof.

\subsection{The tropical moduli space}
\label{sec:tropmodsp}

\begin{figure}[t]
\centering
{\footnotesize
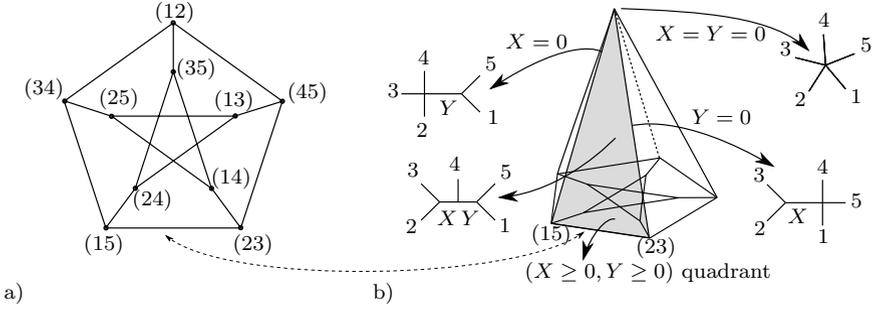}
\caption{a) A slice of the tropical moduli space
  $\cM^{\trop}_{0,5}$. b)~$\cM^{\trop}_{0,5}$, with a specific quadrant in grey. }
\label{fig:M05trop}
\end{figure}

The moduli space $\cM(\Gamma)$ associated to a particular tropical
graph $\Gamma=(G,w,\ell)$ is the cone spanned by the lengths of its
inner edges, modulo the discrete automorphism group of the graph;
\begin{equation}
 \cM(\Gamma) = \IR_+^{|E(G)|}/{\rm Aut}(G)\,.
\label{e:tropmodgr}
\end{equation}
The tropical moduli space of all genus $g$, $n$-punctured graphs is
defined by gluing all these cones together
\cite{Brannetti:2009,Caporaso:2010}, we denote it $\Mt$.
In physical terms, this definition is that of the moduli space of
Feynman or worldline graphs including graphs with counter-terms. We reproduce a few
examples below, and start with $\Mtgn0n$. These latter spaces are themselves
tropical varieties (actually, tropical orbifolds), of dimension $(n-3)$
\cite{Mikhalkin:2006a,Mikhalkin:2007}. Because of the stability
condition \eqref{e:stabcond}, the smallest allowed value of $n$ is
$n=3$. The space $\Mtgn03$ contains only one graph with no modulus (no
inner length): the three-punctured tropical curve.
The space $\Mtgn04$ has more structure; it is isomorphic to the
three-punctured tropical curve and contains combinatorially distinct
graphs which have at most one inner length, as shown below in figure
\ref{fig:m04}.
\begin{figure}[b]
\centering
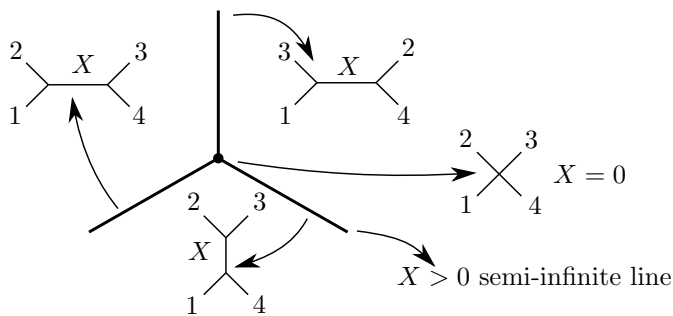
\caption{Thick line; $\cM_{0,4}^{\trop}$. The $X$ coordinate gives the
  length of the inner edge of the various graphs. $X=0$ is common to
  the three branches.}
\label{fig:m04}
\end{figure}
The space $\cM^\trop_{0,5}$ is a two dimensional simplicial complex
with an even richer structure (fig.~\ref{fig:M05trop}).
At genus one, $\Mtgn11$ is also easily described. A genus-one
tropical graph with one leg is either a loop or a vertex of weight
one. Hence, $\Mtgn11$ is isomorphic to the half-infinite line
$\{T\in\IR_+\}$. The graph with $T=0$ is the weight-one vertex, while nonzero $T$'s
correspond to loops of length $T$.

For generic $g$ and $n$, Euler's relation gives that a stable graph
has at most $3g-3+n$ inner edges and has exactly that number if and
only if the graph is pure and possess only trivalent vertices. This
implies that $\Mt$ is of dimension $3g-3+n$ almost everywhere, while 
some of its subsets (faces) are of higher co-dimension.
Finally, note that there also exist a description of $\Mt$ in terms of
the category of ``stacky fans'', discussed in refs.~\cite{Chan:2011,Chan:2012}.

\section{Classical geometry and the tropical limit}
\label{sec:class-geom-trop}

\subsection{Riemann surfaces and their Jacobians}
\label{sec:class-riem-jac}
\begin{figure}[b]
\centering
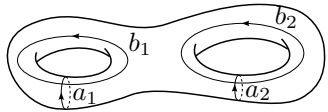
\caption{Canonical homology basis, example for $g=2$.}
\label{fig:canhomolsurf}
\end{figure}

Let $\Sigma$ be a generic Riemann surface of genus $g$ and let
$(a_I,b_J)$, $I,J=1,...,g$ be a canonical homology basis on $\Sigma$
with intersection $a_I\cap b_J=\delta_{IJ}$ and
$a_I\cap a_J= b_I\cap b_J=0$, as in fig.~\ref{fig:canhomolsurf}.

The abelian differentials $\omega_I$, $I=1,...,g$ are
holomorphic 1-forms, they can be normalized along $a$-cycles, so that their
integral along the $b$-cycles defines the period matrix $\bm \Omega$
of $\Sigma$:
\begin{equation}\label{e:norm-holdiff}
 \oint_{a_I} \omega_J = \delta_{IJ}\,, \qquad \oint_{b_I}
\omega_J =
\Omega_{IJ}\,.
\end{equation}
The modular group $\mathrm{Sp}(2g,\IZ)$ at genus $g$ is spanned by the
$2g\times 2g$ matrices of the form $\bigl(\begin{smallmatrix} \bm A&\bm B\\
  \bm C&\bm D \end{smallmatrix} \bigr)$,
where $\bm A,\bm B,\bm C$ and $\bm D$ are $g\times g$ matrices with
integer coefficients satisfying $\bm A \bm B^t = \bm B \bm A^t$,
$\bm C \bm D^t = \bm D \bm C^t$ and
$\bm A\bm D^t-\bm B \bm C^t = \bm{\mathrm{Id}} _g$, with
$\bm{\mathrm{Id}} _g$ the identity matrix. At $g=1$, the modular
group reduces to $\mathrm{SL}(2,\IZ)$. The Siegel upper half-plane
$\cH_g$ is the set of symmetric $g\times g$ complex matrices with
positive definite imaginary part
\begin{equation}
 \cH_g = \{\bm \Omega \in\mathrm{Mat}(g\times g,\IC) : \bm \Omega^t = \bm 
\Omega,
\Im(\bm \Omega)>0\}\,.
\label{e:siegel}
\end{equation}
The modular group $\mathrm{Sp}(2g,\IZ)$ acts on $\mathcal{H}_g$ by
$ \bm{ \Omega} \mapsto (\bm{A \Omega + B}) (\bm{C \Omega+D})^{-1}$.
Period matrices of Riemann surfaces are elements of the Siegel upper
half-plane and the action of the modular group on them is produced by
the so-called Dehn twists of the surface along homology cycles.
The Jacobian variety $J(\Sigma)$ of $\Sigma$ with
period matrix $\bm \Omega$ is the complex torus
\begin{equation}
\label{e:classjac}
J(\Sigma)=\IC^g/(\IZ^g+\bm \Omega \IZ^g)\,.
\end{equation}
Integration along a path $C$ between two points $p_1$ and $p_2$ on the surface
of the holomorphic one-forms defines the classical Abel-Jacobi map 
$\mu$:
\begin{equation}
 \mu(p_1,p_2)= \int_{p_1}^{p_2}{(\omega_1,...,\omega_g)} \mod
\IZ^g+\bm \Omega \IZ^g\,.
\label{e:AbelJacobi}
\end{equation}
As in the tropical case, the right-hand side of \eqref{e:AbelJacobi}
does not depend on the integration path.  Note that, apart for the
very special case of genus one where $\mu(\Sigma_1)\cong \Sigma_1$,
the image of a genus $g\geq2$ Riemann surface $\Sigma_g$ by $\mu$ is
strictly contained in $J(\Sigma_g)$,
$\mu(\Sigma_g)\subsetneq J(\Sigma_g)$.

\subsection{Riemann surfaces and their moduli spaces $\M$, $\Mbar$}
Smooth Riemann surfaces of genus $g$ with $n$ punctures span a moduli
space denoted $\M$ of complex dimension $3g-3+n$ whose coordinates
are called the moduli of the surface. This space is not compact, since
surfaces can develop nodes when non-trivial homotopy cycles pinch off and
give rise to nodal surfaces with double points.  The result of adding
all such nodal curves to $\M$ is the well known Deligne-Mumford
compactified moduli space of curves $\Mbar$ \cite{Deligne:1969}. The
nodal curves are then boundary divisors in $\Mbar$. There exist two
types of such degenerations, called separating and non-separating
degenerations.
\begin{figure}[b]
\centering
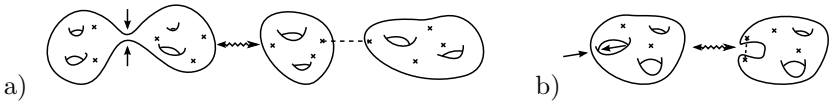
 \caption{a) A separating degeneration. b) A non-separating degeneration. Dashes
represent double points.}
\label{fig:degenMgnbar}
\end{figure}
A separating degeneration splits off the surface into a surface with
two components linked by a double point, while a non-separating
degeneration simply gives rise to a new surface with two points
identified, whose genus is reduced by one unit (see
fig.~\ref{fig:degenMgnbar}). Further, no degeneration is allowed to
give rise to a nodal curve that does not satisfy the stability
criterion shared with tropical graphs \eqref{e:stabcond}. As a
consequence, a maximally degenerated surface is composed of
thrice-punctured spheres.

\begin{figure}[t]
\centering
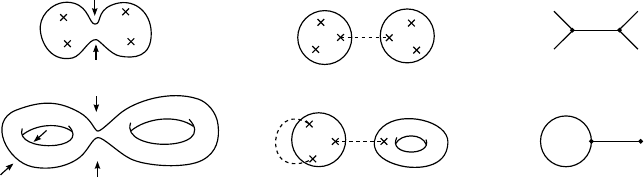
\caption{Leftmost column; degenerating surfaces. Centre;
  nodal curve. Rightmost; dual graphs.}
\label{fig:dualgens}
\end{figure}

These degenerations induce a stratification on $\Mbar$. It is
characterized by the so-called ``dual graphs''. These encore the
combinatorial structure of the nodal curves and the co-dimension of
the boundary divisors. They are defined as follow. Take a nodal curve.
Draw a line that goes through each pinched cycle and turn each
non-degenerated component of genus $g\geq0$ into a vertex of weight
$g$. Draw ``legs'' attached to the graph for each marked point on the
surface. See examples in fig.~\ref{fig:dualgens}.

A surface where a node is developing locally looks like a neck whose
coordinates $x$ and $y$ on each of its side obey the following
equation
\begin{equation}
 xy=t\,,
 \label{e:loccoordneck}
\end{equation}
where the complex number $t$ of modulus $|t|<1$ is a parameter
measuring the deformation of the surface around the boundary divisor
in $\Mbar$. The surface is completely pinched when $t=0$. After a
conformal transformation, this surface is alternatively described by a
tube of length $-\log |t|$ and the tropicalization procedure will turn
these tubes into actual lines.

\subsection{Tropicalizing $\M$}
\label{sec:tropicalizing}
The following schematic construction, not really described explicitly
in the tropical geometry literature, is based on the standard physical
$\apt$ limit of string theory amplitudes. The essential difficulty of
the $\apt$ of string theory is that the objects that we are taking
limits of are integrals over $\M$, which is not a compact space. This
integrand has singularities at the various boundary divisors, and one
is forced to study the integral locally to take the limit.

\paragraph{Decomposition of the moduli space}

We proceed as follows: $\M$ is decomposed into a disjoint union of domains such
that each of them gives rise to a combinatorially distinct set of
tropical graphs;
\begin{equation}
 \M = \bigsqcup_{G}\,\cD_G
 \label{e:Mgndecomp}
\end{equation}
where $\sqcup$ symbolizes disjoint union and in the bulk of each
domain $\cD_G$ lies a nodal curve of $\Mbar$ with dual graph $G$.
The existence of such a decomposition is intuitively clear from the
stratum structure of the moduli space. To obtain a disjoint union as
in eq.~\eqref{e:Mgndecomp}, just ensure to redefine potentially
overlapping domains so as to remove the intersections. This
decomposition is not unique.  The boundaries of the domains can
be deformed so long as they does not start to absorb neighboring
singularities. An explicit decomposition based on minimal area metrics
can be found in Zwiebach's work \cite{Zwiebach:1992ie}, on which we
come back below.

In each of these domains we have local coordinates -- like $t$ in
\eqref{e:loccoordneck} -- that parametrize the surfaces.
Let us exclude the marked points of the following discussion, for
simplicity.
Close to the singularity, the surface is developing a
certain number $N$ of narrow necks or long tubes: as many as there are
inner edges in $G$. Each of them are parametrized by a complex
parameter $t_j$ for $j=1,\ldots,N$ whose collection form a set of local
coordinates.
The tropical graph is obtained by forgetting the phase on the
$t_j$'s. The lengths $T_j$ of its edges are then given by
\begin{equation}
T_j=-\ap\log|t_j|\,. 
\label{e:troplengths}
\end{equation}
Hence, to obtain edges of finite size, the $t_i$'s should actually define
families of curves with a particular scaling, depending on $\ap$,
dictated by \eqref{e:troplengths}:
\begin{equation}
 t_j =  \exp(i(2 \pi \phi + i T_j/\ap)),\quad |t_j| \to0\,,\quad \phi\in[0,2\pi[
\end{equation}
The rest of the $3g-3$ moduli describe the non-degenerating parts of
the surface.
 The field theory limit procedure requires to integrate out
these moduli to create weighted vertices.  Alternatively, keeping
$t_j$ fixed in \eqref{e:troplengths} corresponds to sending $T_{j}$ to
zero, which is consistent with the definition of weighted vertices as
the result of specialized loops. In this paper, we do not describe the
technology to handle these type of integration.\footnote{The
  literature on this is too vast to be summarized here, see however
  recent developments at genus one
  \cite{Angelantonj:2011br,Angelantonj:2015rxa,Angelantonj:2013eja},
  two \cite{D'Hoker:2013eea,D'Hoker:2014gfa,Pioline:2015qha} and
  higher genus \cite{Pioline:2014bra}.}

Two specific kinds of domains are particularly interesting from the
physical perspective that shall be called ``analytic domain'' and
``maximally non-analytic domain{s}'', respectively. This terminology
is borrowed from \cite{Green:1999pv} and refers to the analyticity of
the string amplitudes restricted to these domains. The analytic domain
corresponds to the most superficial strata of $\Mbar$ which
tropicalizes to the $n$-valent weight-$g$ vertex. In this domain, the
string theory integrand has no poles in the $t_j$ moduli and it is
possible to take the limit $\apt$ directly inside the integral. This
gives the primary UV divergences of the field theory amplitudes, at
any loop order, the most divergent parts of field theory amplitudes.
The maximally non-analytic domains correspond to the deepest strata of
$\Mbar$ and give rise to pure tropical graphs made of trivalent
vertices only; this is the field theory unrenormalized amplitude.

\paragraph{Comment on the relation to the minimal area metrics
  formalism.} So far, what was described was a formal
construction. Zwiebach in \cite{Zwiebach:1992ie} defined an explicit
decomposition of $\M$ based on a ``minimal area metrics''
\cite{Zwiebach:1990ew,Wolf:1992bk}, which we summarize now. The idea
is that for any given Riemann surface, there exists a unique metric of
minimal area for which the length of any non-contractible closed loop
is greater than $2\pi$. This metric foliates the surface by closed
loops of length $2\pi$, and Feynman graphs are basically obtained by
drawing on the surface a path that intersect orthogonally these
curves. More precisely, if the height of a local foliation is bigger
than $2\pi$, then it corresponds to a propagator, if no foliation have
height greater than $2\pi$ one is dealing with the genus-$g$ $n$-point
string vertex, etc. (see more details in sec.~6
of~\cite{Zwiebach:1992ie}). Along the time foliation, the local
parameters (now real) presumably give rise to the lengths of the
tropical graphs via the standard scaling \eqref{e:troplengths} in the
\apt~limit.\footnote{As is explained later in
  sec.~\ref{sec:contactterms}, and in the explicit computations in
  sec. \ref{sec:troppf}, here we actually do not need certain domains
  (=vertices) of the string field theory decomposition, those that
  correspond to graphs that contain vertices of weight 0 and valence
  $v\geq4$. They contribute subleading terms in the limit. Therefore,
  an explicit decomposition of the kind we need here could be obtained
  in principle from Zwiebach's by removing the union of all of these
  domains from the decomposition of eq.~\eqref{e:Mgndecomp} and gluing
  them together to form an ``outer'' domain $\cD_{0}$. The
  decomposition then becomes
  $\M=\sqcup_G \mathcal{D}_G \sqcup \mathcal{D}_0$, and the string
  theory integral has no support at leading order over $\cD_0$.}  But
it is not at all obvious that it is doable in practice to implement
this construction in the context of the field theory limit of string
theory which is the one we investigate here. In particular, when
possible (i.e. when there is no ``Schottky problem'', so up to three
loops),\footnote{The Schottky problem is to identify the locus of the
  moduli space of Riemann surfaces (of dimension $3g-3$) inside that of
  Jacobian varieties, of dimension $g(g+1)/2$. These dimensions
  coincide up to three loops, with a subtlety at $g=1$. At $g=4$, the
  problem is solved and the locus is determined by the zero locus of a
  certain modular form called the Schottky-Igusa form.} it is more
convenient to parametrize the moduli space of surfaces in terms of
period matrices. Below we use an such explicit decomposition.

The objective of Zwiebach's construction was to give a set of Feynman
rules to construct formally full string theory amplitudes using
propagators and vertices, in order to obtain a second quantized path
integral formulation of string theory for instance. Therefore, the
consistency of the quantization of his string field theory essentially
guarantees the following. The $\apt$ limit of the string field theory
is a well-defined quantum field theory. Moreover, it could be possible
to extract field theory Feynman rules from the string field theory
ones in this way.\footnote{Actually the bosonic closed string probably does not
  have a naive field theory limit anyway because of the Tachyon.}
This is not the goal that we are pursuing here.

In conclusion of this discussion, as far as computing string
amplitudes an taking their field theory limit is concerned, first
quantization appears to be the most efficient formalism. It is
therefore not in the scope of this paper to investigate further the
analysis of the formal field theory limit of Zwiebach's string\textit{
  field }theory. Instead, we will now expose how to implement the
tropical technology in order to extract field theory limits of string
amplitudes in their explicit and compact first-quantized form.

\paragraph{Classical versus tropical.}
The definitions of previous sections lead to the following three
facts:
\begin{enumerate}[(i)]
\item When going from surfaces to graphs, one-half of the homology
  disappears: the $a$-cycles pinch and the strings become point-like.
\item In particular, since the Abel-Jacobi map maps the $a$-cycles
  to the real part of the Jacobian variety, the imaginary part of the
  period matrices $\Im \bm \Omega$ of tropicalizing surfaces should be
  related to the period matrix of the tropical graph $\bm K$.
\item The classical holomorphic one-forms become one-forms that are
  constant on the edges.
\end{enumerate}
We want to interpret these in the context of the tropical limit.

\begin{figure}[t]
  \begin{picture}(100,40)
\put(-100,0){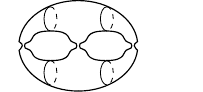}
\put(0,20){$\bm \Omega^{(2)}_{\alpha'}=
\frac 1 {2 i \pi} \begin{pmatrix}
  -\log(t_1 t_3)& \log(t_3)\\ \log(t_3) & 
-\log(t_2 t_3) \end{pmatrix} + O(\alpha',t_i)$}
\end{picture}
\caption{Degenerating Riemann surface parametrized by local coordinates
$t_1,t_2,t_3$ and its period matrix. The $1/(2i\pi)$ normalization follows 
Taniguchi's \cite{Taniguchi:1991} but differs from Fay's in the standard ref
\cite{Fay:1973}~eq.~(54) because of different normalizations (recall 
eq.~\eqref{e:norm-holdiff}).}
\label{fig:loccoordgen2}
\end{figure}

Let us start with period matrices, restricting first to those of 1PI
pure graphs.  Consider a families of curves degenerating towards a
maximal codimension singularity, with local parameters $t_i$, as in
\eqref{e:troplengths}. Taniguchi showed in \cite{Taniguchi:1991} that
the elements of the family of period matrices are given by a certain linear
combination of logarithms of the $t_i$'s, in a rather obvious
combination. An example is shown in fig.~\ref{fig:loccoordgen2}, where
the period matrix \eqref{e:permatrixex1} of the two-loop tropical
graph of fig.~\ref{fig:jacgen2};
$\bm \Omega^{(2)}_\ap = i \bm K^{(2)}/(2\pi\ap) + O(1)$ is immediately
recovered, using the tropical scaling~\eqref{e:troplengths}. This
procedure generalizes straightforwardly to other cases and we obtain
that, in a given domain, the tropicalizing families of curves defined
by \eqref{e:troplengths} have period matrices that approach the period
matrix $\bm K$ of the tropical graph as
\begin{equation}
\Re \bm \Omega_{\ap} = \bm M_0 + O(\ap,t_i)\,,\qquad
 \Im \bm \Omega_{\ap} =\bm  K/(2\pi\ap)+\bm M_1+O(\ap,t_i) \,,\label{e:jacscal}
\end{equation}
where $\bm M_0$ and $\bm M_1$ are constant matrices with real
coefficients. The $(1/2\pi)$ normalization is discussed shortly after
eq.~\eqref{e:riemann-bilin}.  In total, at leading order and up to a
rescaling by $\ap$, the tropical Jacobian is the imaginary part of the
complex one.\footnote{For non pure graphs, one has to be more careful
  with such a statement, see the remark at the end of sec.~\ref{sec:jactrop}.}
To extend this to 1PR graphs, observe that the one-forms have zero
support on the separating edges. In a domain corresponding to a dual
graph $G$ where an edge $e$ splits off $G$ into two 1PI graphs $G_1$
and $G_2$, let $t_e$ be a local coordinate parametrizing such a
separating degeneration. The period matrix of the degenerating curve
is given by;
\begin{equation}
 \bm \Omega^{(t_e)} = \begin{pmatrix}
\bm \Omega_1&0\\
0&\bm \Omega_2
\end{pmatrix}+O(t_e) \,,
\label{e:jacscalpr}
\end{equation}
which can be tropicalized further following the previous discussion
and provides the same splitting for the period matrix of the
corresponding tropical graphs
\begin{equation}
\bm  K = \begin{pmatrix}
\bm K_1&0\\
0&\bm K_2
\end{pmatrix}\,.
\end{equation}

The holomorphic one-forms, at a neck $j$ parametrized by $t_j$, behave
locally as on the cylinder:
\begin{equation}
 \omega_I=\frac c {2 i \pi} \frac{\d z} z + O(t_i)\,,
\label{e:limitomega}
\end{equation}
where $c=1$ or $c=0$ depending on whether the cycle $b_I$ contains
the node $i$ or not. The Abel-Jacobi map \eqref{e:AbelJacobi} then
reduces to 
\begin{equation}
 \int^z \omega_I = \frac c {2 i \pi} \log(z) \in J(\Gamma)
\label{e:limitAJ}
\end{equation}
where it is now clear that the phase of $z$ is mapped to real parts in
$J(\Gamma)$ in the tropical limit. Moreover, consider the following
tropicalizing family of points $z$ on the tube $j$:
\begin{equation}
 z_\ap = e^{i(\theta + i Y/\ap)}
\label{e:ptsscal}
\end{equation}
where $\theta\in[-\pi;\pi[$ and $Y$ is a positive real number. This
yields the tropical limit of the Abel-Jacobi map
\begin{equation}
 2\pi\ap\int^z \omega_i =i\int^Y \omega^\trop_I  = iZ + O(\ap) \in
\ap \Im J(\Sigma_\ap) \equiv J(\Gamma)\,,
\label{e:limitAJtrop}
\end{equation}
where we used that $\omega^\trop_I=1$ on $B_I$. This result is in
accordance with \eqref{e:jacscal}. Finally, these equations are
compatible with Riemann bilinear relations
\begin{equation}
	\int \omega_I\wedge \bar\omega_J = \Im \Omega_{IJ}\,,
\label{e:riemann-bilin}
\end{equation}
which descend to a tropical version (upon multiplication by $\ap$):
\begin{equation}
	\ap \int \omega_I\wedge \bar\omega_J 
	\underset{\apt}{\longrightarrow} \frac{\ap}{(2\pi)^2} 
	\int \frac{d z\wedge \d \bar z}{|z|^2}=\frac{1}{2\pi} \int \d Y = 
\frac{K_{IJ}}{2\pi}
\label{e:riemann-bilin-lim}
\end{equation}
where $Y$ is defined in eq.~\eqref{e:ptsscal}. This eventually
justifies the normalization in eq.~\eqref{e:jacscal}. Another explicit
cross-check of the normalization is provided later at one loop (see
sec.~\ref{sec:one-loop}) where one has to identify the imaginary part
of the modular parameter $\tau$ with a rescaled Schwinger proper time
$T/(2\pi \ap)$. See also the discussion of \cite[pp.~218]{pol1}.

\subsection{The tropical prime form}
\label{sec:tropprimeform}
Let $\Sigma$ be a Riemann surface of genus $g$ with period matrix
$\bm \Omega$.  The classical Riemann theta function is defined on the
Jacobian variety of $\Sigma$ by
\begin{equation}
  \theta(\bm \zeta|\bm \Omega) =
  \sum_{\bm n\in\IZ^g} e^{i\pi \bm n\cdot \bm \Omega\bm  n} 
  e^{2 i \pi \bm n  \cdot \bm \zeta}
\label{eq:theta-def}
\end{equation}
where $\bm \zeta\in J(\Sigma)$ and $\bm \Omega \in\cH_g$. Here and
below we call Fourier expansions these series in
$e^{2i\pi\Omega_{IJ}}$. Theta functions with characteristics are
defined by
\begin{equation}
  \begin{aligned}
    \theta\ab (\bm \zeta|\bm \Omega) = &\,e^{i \pi \bm \beta\cdot \bm
      \Omega \bm \beta + 2 i \pi {\bm \beta\cdot}(\bm \zeta+ \bm \alpha)
    }
    \theta({\bm \zeta} + \bm \Omega \bm \beta+\bm \alpha|\bm \Omega)\\
    =& \sum_{\bm n\in\IZ^g} e^{i\pi (\bm{n}+\bm{\beta})\cdot \bm \Omega (\bm{n}+\bm{\beta})} e^{2 i
      \pi (\bm{n}+\bm{\beta})\cdot (\bm{\zeta}+\bm{\alpha})}
    \label{e:thetachar}
  \end{aligned}
\end{equation}
where $\bm \alpha,\bm \beta\in\frac 12 (\IZ/2\IZ)^{2g}$ are the theta
characteristics. There are $2^{2g}$ of them and the parity of the
scalar product $4\,\bm{\alpha}\cdot\bm{\beta}$ modulo 2 corresponds to
the parity of both the spin structure and the theta function (in $z$);
$\tfrac12(2^{2g}+2^{g})$ are even, the remaining
$\tfrac12(2^{2g}-2^{g})$ are odd.

The prime form \cite{Fay:1973,Mumford:1983}, is an object of central
importance for string
amplitudes~\cite{Verlinde:1986kw,D'Hoker:1988ta}. It is defined by
\begin{equation}
 E:(x,y)\in\Sigma\times \Sigma \longrightarrow E(x,y)=
\frac{\theta\ab(\int_x^y
(\omega_1,...,\omega_g)|\bm{\Omega})}{h\ab(x)h\ab(y)}\,\in 
\IC\,,
\label{e:primeform}
\end{equation}
where $\ab$ is an odd theta characteristic and $h\ab$ are
half-differentials defined on $\Sigma$ by
\begin{equation}
  h\ab(z)^{2}= 
  \sum_{i=1}^g
  \omega_I(z)\partial_I \theta\ab({ 
    0}|\bm{\Omega})\,.
\label{e:halfdiff}
\end{equation}
In this way, the prime form is a differential form of weight
$(-1/2,0)$ in each variables. It is also independent of the spin
structure $\ab$ (this is not obvious from this definition, see for
instance \cite{D'Hoker:1988ta}). In a sense, it generalizes
$(x-y)/\sqrt{\d x}\sqrt{\d y}$ to arbitrary Riemann surfaces and in
particular it vanishes only along the diagonal $x=y$.
It is multi-valued on $\Sigma \times \Sigma$ since it depends on the
path of integration in the argument of the theta function. More
precisely, it is invariant up to a sign if the path of integration is
changed by a cycle $a_I$, but it picks up a multiplicative factor when
changing the path of integration by a cycle $b_J$
\begin{equation}
E(x,y)\to\exp(- \Omega_{JJ}/2 - \int_x^y \omega_J) E(x,y)
\label{e:multE}\,.
\end{equation}

We define the tropical prime form to be the result of the following
limit:
\begin{equation}\label{e:tropE}
E^{\trop}(X,Y) := -\lim_{\apt}\left(
\ap\log{\big|}E(x_\ap,y_\ap|\bm \Omega_\ap){\big|}\right)
\end{equation}
where $\bm \Omega_{\ap}$ are the period matrices of a family of curves
$\Sigma_{\alpha'}$ tropicalizing as in \eqref{e:jacscal} to a graph
$\Gamma$, 
\begin{equation}
  \bm \Omega_{\ap} = i {\bm K}/(2\pi\ap) + \ldots
\label{e:tropscalOmega}
\end{equation}
where the $\ldots$ indicate subleading $\alpha'$ terms and $K$ is the
period matrix of $\Gamma$. The two families of points $x_{\ap},y_\ap$
on $\Sigma_{\ap}$ degenerate as in \eqref{e:ptsscal} to $X$ and $Y$
on $\Gamma$. By the Abel-Jacobi map, we also have a family of elements
in the family of Jacobian
\begin{equation}
  \bm \zeta_{\ap}\in J(\Sigma_{\ap})\,,
\label{eq:degen-jac}
\end{equation}
that degenerates to an element of the tropical Jacobian
\begin{equation}
\bm Z\in J(\Gamma)\,,
\label{eq:degen-jac-trop}
\end{equation}
in such a way that
\begin{equation}
 \bm{\zeta}_\ap = i \bm{Z}/(2\pi\ap)+\ldots
\label{e:tropscalpts}
\end{equation} 
where again the dots indicate subleading terms.

Now comes one of the most important results of this work, the
computation of the field theory or tropical limit of the prime form.
\begin{prop}
 The tropical prime form defined as above corresponds at any loop
order to the graph distance $d_\gamma(X,Y)$ between $X$ and $Y$ along
a path $\gamma$:
\begin{equation}
 E^\trop(X,Y)=d_\gamma(X,Y)\,.
\label{eq:tropEconj}
\end{equation}
\label{prop:tropE}
\end{prop}
\paragraph{Proof.}
The difficult point in this proof lies in the fact that, although the
prime form does not depend on the spin structure, its various
constituents do. We will actually turn this to our advantage and use
lemma.~\ref{lem:dist} to pick an adequate spin structure. More
precisely, having defined (fixed) the families of points $x_\ap,y_\ap$
and their limits $X,Y$ on the graph, there will always exist a class
of convenient spin-structures that make the computation easier.

The first ingredient of the proof is the limit of the theta functions
in the numerator of $E$. Below, we suppress the $\ap$ index but keep
in mind that we deal with families of curves.
Let us first describe the case of theta functions without
characteristics defined in eq.~\eqref{eq:theta-def}. Given the above
scaling, in the series expansion \eqref{eq:theta-def}, all terms but
one are exponentially suppressed:
\begin{equation}
  \label{eq:exple-term}
  e^{i \bm{n\cdot \Omega n}+2i \bm{\zeta\cdot n}}\to 0\,,
\end{equation}
except for $\bm{n}=0$, where we have
$ e^{i \bm{n\cdot \Omega n}+2i \bm{\zeta\cdot n}}=1$.
The case of theta functions with (odd) characteristics is
similar; generic terms in the sum read
\begin{equation}
  \label{eq:generic-theta-char}
  e^{i\pi (\bm{n}+\bm{\beta})\cdot \bm \Omega (\bm{n}+\bm{\beta})} e^{2 i
      \pi (\bm{n}+\bm{\beta})\cdot (\bm{\zeta}+\bm{\alpha})}\,.
\end{equation}
By definition of an odd theta characteristics, $\bm \beta\neq \bm 0$,
and $\bm \beta+\bm n\neq$ for all $\bf n$ since the elements of
$\bm \beta$ are half-integers. Therefore, all terms in the
expansion~\eqref{eq:generic-theta-char} are exponentially suppressed
by the positive-definiteness of $\Im \bf \Omega$.
The leading order term of the theta sum is reached for two values of
$\bm n$,
\begin{equation}
  \label{eq:leading-n}
  \bm{n}=0\ \mathrm{and}\ \bm{n}=-2 \bm{\beta}\,,
\end{equation}
and the leading order asymptotics reads
\begin{equation}
  \label{eq:tropcharlimsum}
  \theta\ab(\bm{\zeta}|\bm{\Omega}) =
  e^{i\pi \bm{\beta}\cdot \bm{\Omega}\bm{\beta}} 
\left(e^{2i\pi(\bm{\zeta}+\bm{\alpha})\cdot\bm{\beta}}+e^{-2i\pi(\bm{\zeta}+\bm{\alpha})\cdot\bm{\beta}}\right)
+\ldots\,.
\end{equation}
This is rewritten
\begin{equation}
  \label{eq:tropcharlim}
  \theta\ab(\bm{\zeta}|\bm{\Omega}) =
  e^{i\pi \bm{\beta}\cdot \bm{\Omega}\bm{\beta}} 
  e^{2i\pi \bm{\beta}\cdot\bm{\alpha}}
2i\sin(2 \pi \bm{\zeta}\cdot\bm{\beta})
+\ldots\,,
\end{equation}
using that
$e^{2i\pi
  \bm{\beta}\cdot\bm{\alpha}}=-e^{-2i\pi\bm{\beta}\cdot\bm{\alpha}}$
since $2\bm \alpha\cdot\bm \beta \equiv 1/2\ (\rm{mod}\ 1)$ for an odd
theta characteristics.  The prefactor
$e^{i\pi \bm{\beta}\cdot \bm{\Omega }\bm{\beta}}$ renders the
right-hand side of \eqref{eq:tropcharlim} exponentially suppressed,
but the presence of the half-differentials in the prime form is going
to compensate this. From their definition \eqref{e:halfdiff}, we see
that the computation of the limit of the $h\ab$'s is very similar to that
of the theta functions; we just have to include a derivative. The
extremizing values of $\bm n$ are still $\bm 0$ and $-2\bm \beta$,
and, as in eq.~\eqref{eq:tropcharlimsum} we have;
\begin{equation}
  \label{eq:hktrop-1}
    h\ab(x)^{2}= 2 i \pi \sum_{J=1}^{g}
    \sum_{\bm{n}=\bm{0},-2\bm{\beta}} \omega_{J}(x)(n_{J}+\beta_{J})
    e^{i\pi (\bm{n}+\bm{\beta})\cdot \bm \Omega (\bm{n}+\bm{\beta})}
    e^{2i\pi  (\bm{n}+\bm{\beta})\cdot \bm{\alpha}}\,,
\end{equation}
at leading order. Actually, only a subset of the $\omega_{J}(x)$'s
contributes to the sum. While the one-forms $\omega_{J}(x)$ for which
the limiting divisor $X$ of the family $x_{\ap}$ belongs to the cycle
tropical $B_{J}$ do contribute, the other all vanish (recall
\eqref{e:limitomega}). If we call $B_{i_{1}},\ldots,B_{i_{k}}$ the set
of these $k$ cycles (there is always at least one cycle),
\eqref{eq:hktrop-1} reduces to;
\begin{equation}
  \label{eq:hktrop}
  \begin{aligned}
    h\ab(x)^{2}&= 2 i \pi \sum_{r=1}^{k}
    \omega_{i_{r}}(x)
    \sum_{\bm{n}=\bm{0},-2\bm{\beta}} 
    (n_{i_{r}}+\beta_{i_{r}})
    e^{i\pi (\bm{n}+\bm{\beta})\cdot \bm \Omega (\bm{n}+\bm{\beta})}
    e^{2i\pi  (\bm{n}+\bm{\beta})\cdot \bm{\alpha}}\\
    &= 4i\pi e^{i\pi \bm{\beta}\cdot \bm{\Omega }\bm{\beta}}  \bm{\beta}\cdot\bm{\omega}(x)\,.
  \end{aligned}
\end{equation}
To obtain the second line, we first used that the exponential of the
quadratic form was independent of $\bm n$ and factored it out.
Then, we simplified as above the induced cosine using $e^{2i\pi
  \bm{\beta}\cdot\bm{\alpha}}=-e^{-2i\pi\bm{\beta}\cdot\bm{\alpha}}$; 
$\cos(2 \pi \bm\alpha\cdot\bm\beta)=1$. Finally, the $r$ summation was
rewritten as a scalar product.

Collecting the previous results in \eqref{eq:tropcharlim} and
\eqref{eq:hktrop}, we obtain the explicit behavior of the prime form;
\begin{equation}
  \label{eq:trop-E-lim}
  -\ap\log{\big|}E(x_\ap,y_\ap|\bm \Omega_\ap){\big|} = 
  -\frac \ap 2 \log\left(
    \frac{\sin(2 \pi \bm\beta\cdot \bm\zeta_{\ap})}
    {
      \sqrt{\bm\omega(x_{\ap})\cdot \bm\beta}
      \sqrt{\bm\omega(y_{\ap})\cdot \bm\beta}}
  \right)
\end{equation}
where we have reintroduced the explicit index $\ap$, and where the
factor of $1/2$ comes from the absolute value on the left-hand side.

Now we set the characteristics $\bm \beta$ as in lemma \ref{lem:dist}.
With the scaling of $\zeta_{\ap}$ in~\eqref{e:tropscalpts} and
lemma~\ref{lem:dist}, the sine function in \eqref{eq:trop-E-lim} becomes
\begin{equation}
  \label{eq:arg-sin}
  \sin(2 i \pi \mathrm{dist}_{\gamma}(X,Y)/\ap)\,,
\end{equation}
whose logarithm gives
\begin{equation}
  \label{eq:log-sin-trop}
 - \frac{2\pi}{\alpha'}  \mathrm{dist}_{\gamma}(X,Y)\,.
\end{equation}

Then we need to deal with the factors of
${\bm{\beta}\cdot\bm{\omega}(x)}$.  With our choice of
characteristics, $\bm{\beta}\cdot\bm{\omega}(x)$ produces at leading
order a positive integer or half-integer, whose explicit determination
is irrelevant here, as it vanishes in the logarithm in~\eqref{e:tropE}
as $\apt$. The only important thing is that this quantity should not
vanish:\footnote{Otherwise one should extract higher order
  terms from the Fourier expansion in the half differentials. A
  similar type of cancellation would occur in the argument of
  $\sin(2\pi \bm{Z}_{\gamma}\cdot\bm{\beta})$ in
  \eqref{eq:tropcharlim}, and presumably the two would cancel out,
  but the author hasn't been able to show this in full generality.}
this is ensured by the following facts
\begin{itemize}
\item[(i)] The first all entries of both vectors are positive,
\item[(ii)] Then,
  $\bm \beta$ is chosen such that its $\mathbb{Z}_{2}$ cycle passes
  through $X$. This implies, as we demonstrated, that at least one
  cycle $B_J$ for which $X\in B_{j}$ has $\beta_{J}=1/2$.
\end{itemize}
Therefore
$ \bm{\beta}\cdot\bm{\omega}(x)\geq \omega_{J}(x) \beta_{J}\simeq1/2
$.

The proposition is finally proven by inserting \eqref{eq:log-sin-trop} in
\eqref{eq:trop-E-lim}.
{ } \hfill $\square$

Higher order terms can sometimes be required to compute the tropical
limit of some amplitudes in string theory. In principle, they can be
extracted following the same recipe. For the amplitudes treated in this
paper, only the leading order contribution described above will be needed.

\section{String theory amplitudes, tropical amplitudes and the
  tropical limit}
\label{sec:ftl}
In the previous sections, we introduced tropical graphs and showed how
they result from the tropicalization of Riemann surfaces. We are now
ready to introduce string theory amplitudes and describe their $\apt$
limit.

\subsection{The tropical limit of string theory}
\label{sec:tropthm}
Let $\Ast gn(X)$ denote a generic $g$-loop $n$-point string theory
scattering amplitude for a scattering process $X$ (we omit the
reference to the scattering process when it is not necessary). In the
Ramond-Neveu-Schwarz (RNS) formalism, the amplitudes are given by
integrals over the supermoduli space of super Riemann surfaces
$\sM g n$ \cite{Friedan:1985ge,D'Hoker:1988ta,Witten:2012bh}.  In
contrast, the pure spinor \cite{Berkovits:2000fe} and Green-Schwarz
formalisms, naturally give integrals over the ordinary moduli space of
Riemann surfaces, $\M$.

In this paper, we restrict ourselves to the study of the string
amplitudes that can be written as integrals $\M$ only, whether they
come from the pure spinor formalism or from a case where the RNS
formalism produces such integrals.\footnote{We postpone to the
  discussion some comments on the recents works of Witten and Donagi,
  where it is argued that, from the supermoduli space perspective this
  would automatically imply a restriction to genus $g<5$.}
Our amplitudes will therefore assume the generic form:
\begin{equation}
\Ast g n  = \int_{\cM_{g,n}}\dmubos\,{\mathcal F}_{g,n}\,.
\label{e:Mgnamp}
\end{equation}
In the RNS formalism, the integrand ${\mathcal F}_{g,n}$ involves a
spin structure sum that accounts for the periodicity of the worldsheet
fermions $\psi^\mu$. In the cases that we deal with explicitly, the sum
will already be done, so we will not be more precise about that.  The
bosonic measure $\dmubos$ is a $(3g-3+n)$-dimensional measure that can
be traded for an integration over the period matrices for $g=1,2,3$,
where there is no Schottky problem;
\begin{equation}
 \d \mu_{bos} = \frac{|\prod_{1\leq I<J\leq g} \d\Omega_{IJ}|^2}{ |\det \Im
\bm \Omega|^{d/2}} \prod_{i=1}^n \d^2 z_i\,,
\label{e:measure}
\end{equation}
where $d$ is the number of space-time non-compact
dimensions.\footnote{This normalization is non-standard, in the sense
  that the invariant measure has an inverse power of $g+1$. From the
  point of view of the field theory limit though, the $d/2$ is more
  natural, therefore we define the measure in this way and absorb a
  compensating factor in the definition of the integrand. Also in all
  explicit examples below, we will have $d=10$.}  The integrand can be
decomposed further and written as
\begin{equation}
 {\mathcal F}_{g,n}={\mathcal W}_{g,n} \exp({\mathcal Q}_{g,n})\,.
 \label{e:cWdef}
\end{equation}
The function ${\mathcal W}_{g,n}$ carries all the information about
the particular scattering process. The factor
$\exp({\mathcal Q}_{g,n})$ is called the Koba-Nielsen factor. It is a
universal factor present in all string theory amplitudes. Its exponent
reads
\begin{equation}
 {\mathcal Q}_{g,n} = \ap \sum_{1\leq i<j\leq n} k_i\cdot k_j\,\cG(z_i,z_j) \,,
\label{e:expfact}
\end{equation}
with $\mathcal G$ the bosonic Green's function~
\cite{Verlinde:1986kw,D'Hoker:1988ta};
\begin{equation}
 \cG(z_1,z_2) = -\frac {1}2 \log\left(|E(z_1,z_2)|^2\right) +\pi
\Im \left(\int_{z_2}^{z_1} {
\omega_I}\right) (\Im\bm \Omega ^{-1})^{IJ} \Im\left(\int_{z_2}^{z_1} { \omega_J}
\right)\,.
\label{e:bosprop}
\end{equation}
Unlike the prime form, $\mathcal G$ is well defined on the surface;
changes in $\log |E|$ as in \eqref{e:multE} are compensated by the
second term in~\eqref{e:bosprop}.

The procedure of section \ref{sec:tropicalizing} is then implemented
as follows.
Take the decomposition
$\M = \big(\bigsqcup_{i=1}^N \cD_G\big) \sqcup \cD_0$ of
sec.~\ref{sec:tropicalizing}.  In the $\apt$ limit of $\Ast g n $, the following two points hold:
\begin{enumerate}[(i)]
\item Integrating over the domain $\cD_0$ produces only subleading
  contributions:
\begin{equation}
 \int_{\cD_0} \d \mu_{bos}\, {\mathcal F}_{g,n} = O(\ap)\,.
\label{eq:cD0}
\end{equation}
We call $\cD_{0}$ the ``outer'' domain.
\item In each domain $\cD_G$, there exist a function $F_{g,n}$ defined
  over $\cM^\trop(\Gamma)$, the moduli space of tropical graphs
  $\Gamma=(G,\ell,w)$ with combinatorial type $G$, such that:
\begin{equation}
 \int_{\cD_G} \dmubos\,{\mathcal F}_{g,n}=
\int_{\cM^{\trop}(\Gamma)}
\dmutrop\,F_{g,n} + O(\ap)\,.
\label{e:bostropdomain}
\end{equation}
The measure is given by
\begin{equation}
\dmutrop :=(2\pi)^{d/2-|E(G)|}\frac{\prod_{i\in E(G)} \d \ell(i)}{ (\det \bm  
K)^{d/2}}\,,
\label{e:tropmeas}
\end{equation}
where $\bm K$ is the period matrix of $\Gamma$.
\end{enumerate}

Compared to Zwiebach's string field theory~\cite{Zwiebach:1992ie}, in
the field theory limit, only massless modes propagate along edges of
finite lengths. The contribution of massive modes stay localized on
vertices with weights. We shall see this explicitly in the examples
below.

As far as the explicit computations of this paper are concerned, we
will build by hand these decompositions.

Physically, the right-hand side of \eqref{e:bostropdomain} is the
contribution of the Feynman diagrams of field theory in the tropical
representation corresponding to the graph $G$. As above, the integrand
$F_{g,n}$ can be factorized
\begin{equation}
 F_{g,n}=W_{g,n}\exp(Q_{g,n})\,,
 \label{e:Fgn}
\end{equation}
where $W_{g,n}$ and $Q_{g,n}$ descend from their string theory
ancestors. Computing their explicit form gives the tropical
representation of the integrand and is the second step of the
procedure. The extraction of $W_{g,n}$ is straightforward in the cases
of maximal supergravity four-graviton amplitudes discussed later for
$g=0,1,2$ but it is much more intricate in the general case. It
requires in particular to deal with Fourier expansions in
higher genus, and this will not be covered in this paper, although in
principle the procedure of sec.~\ref{sec:tropprimeform} gives a
prescription to extract these terms. As we mentioned already, this
process at genus one is fully understood since the works of Bern
and Kosower \BKcite.

On the other hand, $Q_{g,n}$ is a universal factor and is obtained
from \eqref{e:expfact} by computing the tropical limit of the Green's
function $\cG$, to which we turn now. We have already studied the
limits of both the prime form in \eqref{e:tropE} and the holomorphic
differentials \eqref{e:limitAJ}, therefore all we have to do is to
piece these up to obtain the {tropical Green's function};
\begin{equation}
  \begin{aligned}
\lim_{\apt} \ap \mathcal G(z_1,z_2)&= -\frac {1}2
E^\trop(Z_1,Z_2) -\frac {1}2
\bigg(\int_{Z_2}^{Z_1} {  \omega^\trop_{I}}\bigg) (\bm  K^{-1})^{IJ}
\bigg(\int_{Z_2}^{Z_1} {  \omega^\trop_{J}} \bigg)\\
&:=  G^\trop(Z_1,Z_2)\,.
  \label{e:bostropprop}
  \end{aligned}
\end{equation}
The limit is to be understood as in section \ref{sec:tropprimeform}
and factors of $(2\pi)$ have been consistently reabsorbed in $\omega$
and $\bm \Omega$ to produce $\omega^\trop$ and $\bm K$.
This tropical Green's function coincides with the worldline Green's
function computed directly in \cite{Dai:2006vj} (see also
\cite{Strassler:1992zr,Schmidt:1994zj,Roland:1996np,Schubert:2001he} for earlier
works). Contrary to the tropical prime form, $G^\trop$ is always
independent of the integration path. It follows from these definitions
that the tropical representation of exponential factor in
\eqref{e:cWdef} is given by
\begin{equation}
Q_{g,n}=\sum k_i\cdot k_j G^\trop(Z_i,Z_j)
\label{e:expfactlim}
\end{equation}
We can now collect  \eqref{e:tropmeas} and
\eqref{e:expfactlim} to obtain the following formula;
the tropical representation of \eqref{e:bostropdomain} is
\begin{equation}
  \int \prod_{i\in E(G)} \d \ell(i)\frac{W_{g,n}\,\exp(Q_{g,n})}
  { (\det\bm K)^{d/2}}\,,
\label{prop:sym}
\end{equation}
up to an overall numerical factor of the form $(2\pi)^{m}$. In this
form, $\det(\bm K)$ and $\exp(Q_{g,n})$ are respectively the first and
second Symanzik polynomials obtained from Feynman rules in field
theory,\footnote{There is a slight difference of normalization
  compared to the usual definition given for instance in the classic
  reference \cite{Itzykson-Zuber} where the first and second Symanzik
  polynomials, denoted $\mathcal U$ and ${\mathcal F}$, are related to
  ours by:
  $ \mathcal U = \det K,\, {\mathcal F} = \exp(Q_{g,n}) \det K\,, $
  and where also $\exp(Q_{g,n})$ should strictly speaking be replaced
  by the result of integrating out a global scale factor for the
  lengths of the edges of the graph to go from Schwinger proper times
  to Feynman parameters.}  and $W_{g,n}$ is the numerator of the
Feynman graph integrand. This assertion is physically clear, however,
a direct proof using graph theory would be of interest concerning more
formal aspects of the study of Feynman diagrams.\footnote{Note also
  that in this representation, it is obvious that the first Symanzik
  polynomial does not depend on the positions of the punctures.}
Examples in genus one and two are given in sec.~\ref{sec:troppf}.

We can now phrase the standard $\apt$ limit in the tropical language;
\begin{conj}
  The $\apt$ limit of the string theory integral over $\M$ is given by
  an integral over $\Mt$
\begin{equation}
 \int_{\cM_{g,n}} \dmubos\,{\mathcal F}_{g,n}=
\int_{\cM_{g,n}^{trop}}
\dmutrop\,F_{g,n} + O(\ap)\,,
\label{e:bostrop}
\end{equation}
where
\begin{equation}
 \int_{\cM_{g,n}^{trop}}\dmutrop := \sum_{\Gamma}
\int_{\cM(\Gamma)} \d \mu^\trop\,.
\label{e:tropmeastot}
\end{equation}
The discrete finite sum runs over all the combinatorially distinct
graphs $\Gamma$ of genus $g$ with $n$ legs.  Moreover, the
right-hand side of \eqref{e:bostrop} corresponds to the field theory
amplitude {renormalized} in the scheme induced by string theory. This
scheme is {defined} such that
\begin{equation}
\A g n _\trop=\int_{\cM_{g,n}^{trop}}
\dmutrop\,F_{g,n}
\label{e:AST}
\end{equation}
where $\A g n _\trop$ is the field theory amplitude written in its tropical
representation (in short tropical amplitude) obtained in the field theory limit. 
\end{conj}

The conjecture can be shown in the cases where one starts from a known
string amplitude, mostly because an explicit $\mathcal{F}_{g,n}$ is
needed. In this way, re-expressing the existing tree-level and
one-loop computations in the tropical language, as we do later, can be
considered as a proof of various instances of the conjecture.

\subsection{Counter-terms, contact terms.}

\paragraph{Analytic and non-analytic terms.}
\label{sec:domains}
For simplicity, let us exclude the punctures of the discussion. The
analytic and maximally non-analytic domains have been defined in
section \ref{sec:tropicalizing} by the requirement that the first
should correspond to the more superficial stratum of $\Mgbar_ g $ and
the second should correspond to the deepest strata of $\Mgbar_g$.

In other words, the analytic domain is defined by removing all
neighborhood around the singularities of $\cM_g$.  Therefore it is a
compact space. Inside that domain, the string integrand has no
singularity and the limit may be safely taken directly; the factor
$\ap$ present in the definition of ${\mathcal Q}_{g,n}$ simply sends
$\exp({\mathcal Q}_{g,n})$ to $1$. 
Moreover, the dual graph of the analytic domain is a single vertex of
weight $g$. Physically, such graphs are counter-terms to primary UV
divergences, so this is consistent with the fact these correspond to
the string integral over the analytic domain, as illustrated later in
the one-loop example of section \ref{sec:one-loop}.

The maximally non-analytic domains provide the contributions of the
pure tropical graphs, the worldline graphs made of trivalent vertices
only (graphs with no counter-terms). Summed over, they give the
unrenormalized field theory amplitude, with all of its divergences. We
present in sec.~\ref{sec:two-loop} a computation of a tropical integrand
at genus two in such a domain.

\paragraph{A remark on contact terms}
\label{sec:contactterms}

Feynman rules in non-abelian gauge theories or gravity naturally use
vertices of valency higher than three to implement gauge invariance.
The way that these arise in string theory is different. What is called
a ``contact-term'' in string theory is usually the vertex that results
from integrating out the length dependence of a separating edge in a 1PR
graph, as in \eqref{e:contact-term} below.
\begin{equation}
	\int \Big(
	\centering{ \begin{fmffile}{contact-term-pole}
\fmfsettings
		\fmfframe(0,0)(-4,-14){ \begin{fmfgraph*}(48,36)
		\fmfleft{g1}
		\fmfright{g4}
		\fmf{phantom,tension=1}{g1,v1}
		\fmf{phantom,tension=1}{g4,v2}
		\fmf{plain,tension=0.5,label=$X$}{v1,v2}
		\fmfblob{0.3w}{v2}
		\fmfblob{0.3w}{v1}
      \end{fmfgraph*} }
      \end{fmffile} } \Big)
      \d X =c_0 \times 
		{ \begin{fmffile}{contact-term}
\fmfsettings
		\fmfframe(-6,0)(-4,-14){ \begin{fmfgraph*}(48,36)
		\fmfleft{g1}
		\fmfright{g4}
		\fmf{phantom,tension=1}{g1,v1}
		\fmf{phantom,tension=1}{g4,v2}
		\fmf{plain,tension=2}{v1,i1,v2}
		\fmfdot{i1}
		\fmfblob{0.3w}{v2}
		\fmfblob{0.3w}{v1}
      \end{fmfgraph*} }
      \end{fmffile} }
\label{e:contact-term}
\end{equation}
These integrations are trivial since they are of the form
$\int_{0}^{\infty}\exp{(-s X)}\d X$ where $s$ is a kinematic
invariant. However, prior to any of these trivial integrations, the
locus $X=0$ corresponds geometrically to a lower codimension face in
$\Mt$ and does not carry any localized contribution, it is only after
integration that a contact term is produced.

\paragraph{Maximal simplicity of maximally supersymmetric
  numerators}
\label{sec:maxim-simpl-maxim}
A final note in this section concerns the simplicity of the extraction
of $W_{g,n}$ in the non-analytic regions. Generic string theory models
exhibit chiral ``tachyon poles'', of the form $q^{-1}$ or $q^{-1/2}$
at $g=1$ and generalization thereof at higher genus (see for instance
\cite{Tourkine:2012ip} at $g=2$ in CHL models).
These poles ``soak up'' powers of $\partial G^\trop$ from the
numerators as they extract residues of the form
$\mathcal{W}_{g,n}\exp(\mathcal{Q}_{g,n})|_{q}$ in the Fourier
expansion. This decreases the degree of the loop momentum numerator
polynomials, thereby enforcing supersymmetric cancellations.  The
Bern-Kosower rules were a systematization of this residue extraction
at one-loop, and one of the longer term goal of this tropical limit
project is to extend these rules to higher loops.

In the case of maximally supersymmetric amplitudes, these tachyon
poles are canceled directly at the level of the spin-structure sum and
the technology presented here is usable straight away to extract the
field theory numerators in the tropical or Schwinger proper-time form.
We give an illustration of this at $g=2$ in sec.~\ref{sec:two-loop}
and in the conclusion mention some work in progress at $g=3$ based on
\cite{Gomez:2013sla}.

\section{Explicit computations}
\label{sec:troppf}
In this section, we first review some examples of field theory limits
at tree-level and one-loop which we formulate in the tropical
framework. Then at two loops, we derive the worldline representation
of the four-graviton amplitude in the non-analytic domain from the
full string theory amplitude of D'Hoker \& Phong. We also comment on
UV divergences and counter-terms.
\subsection{Tree level (review)}
\label{sec:trees}
As a warm-up, we start with tree-level scattering amplitudes in string
theory, as was done by Scherk in the early days of string
theory\cite{Scherk:1971xy}. We first look at the simplest example, the
four-tachyon scattering in the bosonic string, then we describe the
case of four-graviton scattering in the type II superstring. The
general case of $n$-particle scattering follows from the same method
as the one reviewed here.

A closed string theory tree-level $n$-point amplitude can be written in the
general form\footnote{We follow the conventions of \cite{BLT}.}:
\begin{equation}
  \Ast0 n =g_c^{n-2} \frac{8 \pi }{\ap }\int_{\cM_{0,n}} \prod_{i=3}^{n-1}
\d^2 z_i\, \langle (c\bar c V_1) (c\bar c V_2) V_3...V_{n-1} (c\bar c
V_n)\rangle\,,
\label{e:ntree}
\end{equation}
where $\d^2 z := \d z \d\bar z$ and $g_c$ is the string coupling
constant. The vertex operators $V_i$ insert the external scattered
states at position $z_i$ on the worldsheet. They depend on the momenta
$k_i$ and possible polarizations $\epsilon_{i}$ of the particles. The
integration over the points $z_1,~z_2$ and $z_n$ is suppressed and
exchanged by the insertion of $c \bar c$ ghosts to account for the
factorization of the infinite volume of the $SL(2,\IC)$ conformal
group.  The integral over the set of $n-3$ distinct complex variables
$z_3,\ldots,z_{n-1}$ spans the moduli space of $n$-punctured genus zero
surfaces $\cM_{0,n}$. The correlation function \eqref{e:ntree} is
computed using Wick's theorem and the correlators
\begin{equation}
\langle X(z,\bar z) X(w,\bar w) \rangle=\mathcal G(z,w) = -\frac{\alpha'}2
\log(|z-w|^2)\,,\quad
\langle c(z) c(w) \rangle =z-w\,,
\label{e:2ptsphere}
\end{equation}
 The ghost correlator is given by
\begin{equation}
 |\langle c(z_1) c(z_2) c(z_n)\rangle|^2 = |z_{12} z_{2n} z_{n1}|^2\,.
\label{e:cghost}
\end{equation}
The correlation function \eqref{e:ntree} can be written as in
\eqref{e:Mgnamp} by defining $\dmubos=\prod_{i=3}^{n-1} \d^2 z_i$ and
\begin{eqnarray}
&& {\mathcal F}_{0,n}:=g_c^{n-2} \frac{8 \pi}{\ap } {\mathcal 
W}_{0,n}(z_{jk}^{-1},\bar
z_{lm}^{-1})\exp( {\mathcal Q}_{0,n})\,,\\
&&{\mathcal Q}_{0,n}:=\ap\sum_{3\leq i <j \leq n-1} k_i\cdot k_j 
\log|z_i-z_j|\,,
\label{e:npttree}
\end{eqnarray}
where $1\leq j,k,l,m\leq n$ and ${\mathcal W}_{0,n}=1$ for the
scattering of $n$ tachyons, while it is a rational function of the
$z_{jk}$ in the general case of massless states scattering. Its
coefficients are made of factors of $\ap$, scalar products of
polarization tensors and external momenta and include the color
structure for gauge theory interactions.

\begin{figure}[t]
\centering
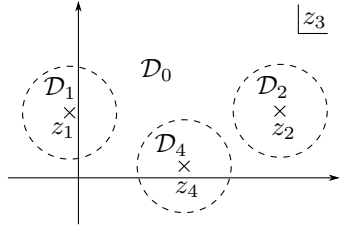
\caption{Decomposition of the moduli space $\mathcal{M}_{0,4}$.}
\label{fig:domains4ptstree}
\end{figure}

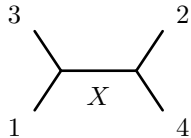
\begin{figure}[b]
\centering
\parbox{40pt}{ \begin{fmffile}{streejac}
\fmfframe(0,10)(0,0){ \begin{fmfgraph*}(60,30)
	\fmflabel{$1$}{g1}
	\fmflabel{$3$}{g2}
	\fmflabel{$2$}{g3}
	\fmflabel{$4$}{g4}
	\fmfleft{g1,g2}
	\fmfright{g4,g3}
	\fmf{plain}{g1,v12,g2}
	\fmf{plain}{g3,v34,g4}
	\fmf{plain,tension=0.70,label=$X$}{v12,v34}
\end{fmfgraph*} }
\end{fmffile} }
\caption{$X$ is the modulus of the tropical graph. The larger it is, the 
closer
$z_1$ from $z_3$.}
\label{fig:4pttree}
\end{figure}

Let us start with the scattering of four tachyon states. The
vertex operator of a tachyon with momentum $k_i$
($k_i^2=-m_\tach^2:=4/\ap$) is a plane wave $ V_j=e^{ik_j\cdot X(z_j,\bar z_j)}$. From
\eqref{e:ntree} we obtain
\begin{equation}
  \begin{aligned}
    \Ast 0 {\mathrm{4-tachyons}} &=
    g_\tach^2\,|z_{12}z_{24}z_{41}|^2 \times \\ &\int \d^2 z_3\, e^{\left(\ap
        k_1\cdot k_3 \log|z_{13}z_{24}|+\ap k_2\cdot k_3 \log|z_{23}z_{14}|+\ap
        k_4\cdot k_3 \log|z_{12}z_{34}|\right)}\,,
  \end{aligned}
\label{e:troptree4}
\end{equation}
where we have introduced the tachyon coupling constant
${g_\tach=8 \pi g_c/\ap}$ and kept $z_1$, $z_2$ and $z_4$ fixed but arbitrary.
Momentum conservation
imposes
$k_1+k_2+k_3+k_4=0$ and the Mandelstam kinematic invariants $s,t,u$ are defined
by
$s=-(k_1+k_2)^2$, $t=-(k_1+k_4)^2,\ u=-(k_1+k_3)^2$. Their sum is the sum of
the squared masses of the particles $s+t+u=\sum_1^4 m_i^2$.
The integral \eqref{e:troptree4} can be computed explicitly and reads
\begin{equation}
  \Ast 0 {4\mathrm{-tachyons}} =2 \pi g_\tach^2
  \frac
  {\Gamma(\alpha(s))\Gamma(\alpha(t))\Gamma(\alpha(u))}
  {\Gamma(\alpha(t)+\alpha(u))\Gamma(\alpha(u)+\alpha(s))\Gamma(\alpha(s)+\alpha(t)
    )}
\label{e:4treelambda}
\end{equation}
where $\alpha(s):=-1-s\,\ap/4$. It has poles in the tachyon kinematic
channels, for instance
\begin{equation}
\Ast 0 {4\mathrm{-tachyons}} \underset{s\to -4/\ap}{\sim} g_\tach^2 \frac{1}
{-s-4/\ap}\,.
\label{e:tachpole}
\end{equation}
We want to recover these poles in the point-like limit in a tropical
language. Physically, these poles originate from regions where vertex
operators collide to one another. Since at tree level in field theory,
there are only poles, the domains $\cD$ of the decomposition in
eq.~\eqref{e:Mgndecomp} precisely correspond to these regions.  At
four points, only one coordinate is free and the domains are just open
discs of radius $\ell$ centered around $z_1$ $z_2$ and $z_4$ called
$\cD_1$, $\cD_2$ and $\cD_4$ as shown in
fig.~\ref{fig:domains4ptstree} (see for instance the classic reference
\cite{pol1}):
\begin{equation}
 \cM_{0,4} = \left(\cD_1\sqcup\cD_2\sqcup\cD_4\right)\sqcup \cD_0\,.
\label{e:dec04}
\end{equation}
We review below how the integrals over each domain provide the $u$,
$t$ and $s$ channel tachyon exchanges, respectively, while the
integral over $\cD_0$ gives a subleading contribution.
We start with the integral over $\cD_1$. As the domains are disjoint,
we have $|z_{21}|>\ell$ and $|z_{41}|>\ell$. Thus, the terms
$\ap k_2\cdot k_3 \log|z_{32}z_{14}|+\ap k_4\cdot k_3
\log|z_{34}z_{12}|$ in eq.~\eqref{e:troptree4} behave like
\begin{equation}
  (-\ap k_1\cdot k_3-4) \log|z_{12}z_{14}|+O(\ap z_{31},\ap \bar
z_{31}) 
\end{equation}
which gives in the integral:
\begin{equation}
\int_{\cD_{1}} \d^2 z_3\frac {|z_{24}|^2}{
|z_{12}z_{14}|^2}\, e^{\ap k_1\cdot k_3 \log \big|\frac{z_{31}z_{24}}{ z_{12}
z_{14}}\big|} + O(\ap)\,,
\end{equation}
The integration over the phase of $z_{31}$ is now trivial, hence we
may change variables to the tropical variable $X$ as in
\eqref{e:ptsscal};
\begin{equation}
c\times z_{31} = \exp({-X / \ap}+i \theta)\,,
\label{e:tropchange4pt}
\end{equation}
where $c$ is a conformal factor given by $c=z_{24} /(z_{12}z_{14})$
and $\theta$ is the irrelevant phase.  In this variable, the closer
$z_3$ is from $z_1$, the larger $X$ is.  The integration measure
becomes $|c|^2 \d^2 z_3 = -\frac{2}{ \ap}\, e^{-2X/\ap} dX\, d\theta$
and the radial integration domain is now
$X\in [-\ap \log \ell,\, + \infty [$.
We integrate out $\theta$, drop the $\ell$-dependent terms, since
they are subleading, and obtain the following contribution to the
amplitude
\begin{equation}
\Ast 0{4-\mathrm{tachyons}}|_{u-\mathrm{channel}} = g_\tach^2 \left(\int
_{0}^\infty dX\
e^{-((k_1+k_3)^2+m_\tach^2)
X} +
O(\alpha')\right)\,.
\label{e:Xtrop4}
\end{equation}
This is simply the exponentiated the Feynman propagator of a scalar
$\phi^3$ theory with coupling constant $g_\tach$ and mass
$m_\tach$. In this form, the modulus $X$ of the graph corresponds to
the Schwinger proper time of the exchanged particle, as in
fig.~\ref{fig:4pttree}.
The same computation can be repeated in the other two kinematic
regions to obtain $s$- and $t$-channel exchanges.
To conclude, one has to check that the integral over $\cD_0$ does
yield only $O(\ap)$ contributions. In the case of tachyon scattering,
this is actually not true, due to the fact that the tachyon acquires
an infinite negative mass squared $m_\tach^{2}=-4/\alpha'$ when
$\apt$, which cancels the exponential damping induced by the factor
$\ap$ already present in $\mathcal{Q}_{0,n}$. This is not surprising
because tachyons generically lead to inconsistencies of the field
theory. In the case of gravitons that we consider next, the limit will
be well-defined and the integral over $\cD_0$ will vanish.

Let us turn to graviton scattering in superstring theory.  The
decomposition remains unchanged. The qualitative difference with the
scalar case is due to the appearance of a non-trivial ${\mathcal W}$.
We will work in a representation of the integrand where all double
poles have been integrated out by parts -- this can always been done.
\footnote{see \BKcite for a one-loop proof and the more recent works
\cite{Mafra:2010jq,Mafra:2011kj,Mafra:2011nv,Mafra:2011nw} for an
extensive study of the tree-level integrand representations, using
integration by parts and fraction by part identities} The tree-level
four-graviton amplitude is written as
\begin{equation}\label{e:4gravtreecf}
\Ast 0{4-\mathrm{grav.}}= \frac{8 \pi g_c^2}{\ap} \langle c\bar c
V_{(-1,-1)}(z_1) c\bar c
V(z_2)_{(-1,-1)} V_{(0,0)}(z_3) c\bar c V_{(0,0)}(z_4)\rangle\,.
\end{equation}
The graviton vertex operators in the $(-1,-1)$ and $(0,0)$ pictures read
\begin{equation}
\begin{aligned}
  &V_{(-1,-1)}(z)= \epsilon_{\mu\nu} \psi^\mu\bar \psi^\nu
  e^{-\phi-\bar \phi}
  e^{ik\cdot X(z,\bar z)}\,, 
  \\
  &V_{(0,0)}(z)=\frac{2 }{ \ap} \epsilon_{\mu\nu} \left(i\bar \partial
    X^\mu+\frac \ap 2 k\cdot\bar \psi \bar
    \psi^\mu\right)\left(i \partial X^\mu+\frac \ap 2 k\cdot\psi
    \psi^\mu\right)e^{ik\cdot X(z,\bar z)}\,.
  \label{e:grav00}
\end{aligned}
\end{equation}
in terms of the polarization tensors
$\epsilon_{\mu\nu}:= \epsilon_\mu \tilde \epsilon_\nu$. The bosonized
superconformal ghost two-point function reads
$\langle\phi(z) \phi(w)\rangle = -\log(z-w)$ while the one of the
fermions reads $\psi^\mu(z)\psi^\nu(w) = \eta^{\mu\nu}/(z-w)$. In
terms of these, the amplitude \eqref{e:4gravtree} can be computed
explicitly (see the classic reference \cite{GSW1});
\begin{equation}\label{e:4gravtree}
\Ast 0{4-\mathrm{grav.}}= \frac{8 \pi g_c^2 }{ \ap} C(s,t,u)\, \mathcal{R}^4\,,
\end{equation}
where $\mathcal{R}^{4}$ is a particular tensorial combination of four
powers of the linearized Weyl tensor
$R^{\mu \nu \rho \sigma} = F^{\mu \nu} \tilde F^{\rho \sigma}$ written
in term the famous tensor $t_{8}$ as $\mathcal{R}^{4}=t_8t_{8}R^{4}$. The
tensors $F$ and $\tilde F$ are on-shell linearized field strengths;
the graviton $i$ with polarization
$\epsilon_i^{\mu\nu}=\epsilon_i^{\mu}\tilde \epsilon_i^{\nu}$ and
momentum $k_i$ has $F_i^{\mu \nu} = \epsilon_i^{[\mu}k_i^{\nu]}$ and
$\tilde F_i^{\rho \sigma} = \tilde
\epsilon_{i}^{[\rho}k_{i}^{\sigma]}$.
The function $C$ and the tensor $t_{8}$ are defined in \cite{GSW1}, we
reproduce them here:
\begin{subequations}
	\begin{align}
 C(s,t,u)&=-\pi\frac{ \Gamma(-\ap s/4)\Gamma(-\ap t/4)\Gamma(-\ap
u/4)}{ \Gamma(1+\ap s/4)\Gamma(1+\ap t/4)\Gamma(1+\ap u/4)}
\label{e:cstu}\,,\\ 
t_8 F^4&=-st (\epsilon_1\cdot\epsilon_3)(\epsilon_2\cdot\epsilon_4)+2t
(\epsilon_2\cdot k_1\, \epsilon_4\cdot k_3\, \epsilon_3\cdot\epsilon_1 
+\epsilon_3\cdot k_4\,
\epsilon_1\cdot k_2\, \epsilon_2\cdot\epsilon_4\nn\\ 
+&\epsilon_2\cdot k_4\, \epsilon_1\cdot k_3\,
\epsilon_3\cdot\epsilon_4 +\epsilon_3\cdot k_1\, \epsilon_4\cdot k_2\,
\epsilon_2\cdot\epsilon_1)+(2\leftrightarrow3) + (3\leftrightarrow4)\,.
\label{e:t8F4}
	\end{align}
\end{subequations}
Schematically, $t_8F^4$ is a polynomial in the kinematic invariants with
coefficients made of scalar products between polarizations and momenta
\begin{equation}
 t_8F^4=C_{s} s + C_t t + C_u u + C_{st} st + C_{tu} tu + C_{us} us\,.
\label{e:t8F4C}
\end{equation}
Since $C(s,t,u)\sim 1/(\ap^3 stu)$, using multiple times the on-shell condition
$s+t+u=0$, the amplitude \eqref{e:4gravtree} can be written as
\begin{equation}
 \Ast 0 4  \sim  \frac {A_s} s + \frac {A_t}t +\frac {A_u}u + A_{0}+O(\ap)
\label{e:A4schem}
\end{equation}
where the $A$'s are sums of terms like $C_s \overline{C}_t$, etc. As the
tensorial structure of this object is rather complicated, we will only focus
ourselves on one particular term; a contribution to $A_u$.
In the correlation function \eqref{e:4gravtreecf}, such a contribution comes
from the following term:
\begin{equation}
  \begin{aligned}
    -(\ap/2)^2(\epsilon_2\cdot\epsilon_4)\tfrac{1}{ z_{24}^2}
    (\epsilon_1\cdot k_4)(\epsilon_3\cdot k_2)\left(\left(\tfrac{1}{
          z_{14}}-\tfrac{1}{ z_{13}}\right)\left(\tfrac{1}{
          z_{32}}-\tfrac{1 }{
          z_{31}}\right)+\tfrac{1}{ z_{13}^2}\right)\times\\
    (-1)(\ap/2)^2\left(\tilde\epsilon_2
      . \tilde\epsilon_4\right)\tfrac{1 }{ \bar z_{24}^2} (\tilde
    \epsilon_1\cdot k_2)(\tilde \epsilon_3\cdot
    k_4)\left(\left(\tfrac{1 }{ \bar z_{12}}-\tfrac{1 }{ \bar
          z_{13}}\right) \left(\tfrac{1 }{ \bar z_{34}}-\tfrac{1 }{
          \bar z_{31}}\right)+\tfrac{1}{\bar z_{13}^2}\right)\,,
  \end{aligned}
\label{e:uchanterm}
\end{equation}
where we have used the conservation of momentum $k_1+k_2+k_3+k_4=0$,
the on-shell condition $\epsilon_i\cdot k_i=0$. It is now
straightforward to check that the term corresponding to $1/|z_{31}|^2$
in the previous expression is accompanied with a factor of
$|z_{12} z_{24} z_{41}|^{-2}$ which combines with the conformal factor
from the $c\bar c$ ghosts integration \eqref{e:cghost} to give
\begin{equation}
-\left(\frac{\ap }{ 2}\right)^3 \int \d^2 z_{31} \frac 1 {|z_{31}|^2} e^{\ap
k_1\cdot k_3 \log|z_{31}|}+O(\ap)\,.
\end{equation}
The phase dependence of the integral is either pushed to $O(\ap)$ terms or
canceled due to level matching in the vicinity of $z_1$. Thus, we can integrate
it out and recast the integral in its tropical form using
the same change of variables as in \eqref{e:tropchange4pt} and one gets the
following contribution to the amplitude of eq.~\eqref{e:4gravtreecf}:
\begin{equation}
 4\kappa_d^2 \left(\int_0^\infty \d X e^{-u X} + O(\ap)\right)\,,
\end{equation}
where $\kappa_d=2\pi g_c$ is the $d$-dimensional coupling constant
that appears in the Einstein-Hilbert action. Other terms are generated
in the exact same manner, by combinations of various massless poles
(even $A_0$, despite that it has no explicit pole structure). The full
amplitude is finally rewritten as an integral over $\Mtgn04$ as
follows;
\begin{equation}
\Ast 0{4-\mathrm{grav.}} \to \A 0 {4-\mathrm{grav.}}= \int_{\Mtgn04}
\dmutrop\, F_{0,4}\,,
\end{equation}
where the measure pulls back to regular integration measure $dX$ on
each edge, and
$F_{0,4}=4\kappa_{d}^{2}t_{8}t_{8}R^{4}\exp\left(-X((k_i+k_3)^2)\right)$
where $i=1,2,4$, depending on the edge of $\Mtgn04$ considered.ft

The generalization to $n$ points is conceptually straightforward,
though combinatorially more involved. The trees with edges of finite
lengths will be generated by similar regions of the moduli space where
the points $z_i$ collides towards one another. Writing out explicitly
this decomposition would not bring any new insight, so we
shall turn to loops now.

\subsection{One loop (review)}
\label{sec:one-loop}
The technical aspects of the point-like limit of one-loop open and
closed string theory amplitudes are well understood. In this review
section, we simply recast in the tropical framework some of the older
results on the subject.
We first focus on the four-graviton type II superstring amplitude
since we are ultimately interested in higher genus four-graviton
amplitudes. 
That amplitude is a nice toy model to see how the tropical limiting
procedure naturally generates the so-called analytic and non-analytic
terms \cite{Green:1999pv,Green:2008uj,Green:2010sp,Green:2010wi} of
the amplitudes together with the counter-terms.
Then we discuss the $n$-point case. We make connection with
the previous section and describe the regions of the string theory
moduli space integral give rise to trees attached to the loop,
recapitulating the Bern-Kosower rules.
\begin{figure}[t]
\centering
 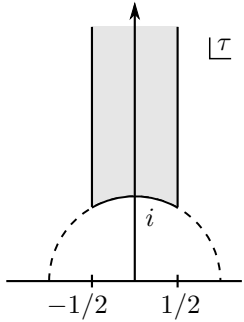
\caption{An $SL(2,\IZ)$ fundamental domain for complex tori.}
\label{fig:fundtorus}
\end{figure}

Let us first review some elements about genus one Riemann surfaces or
elliptic curves. They are complex tori ${\IC/(\IZ+\tau\IZ)}$
parametrized by a complex modulus $\tau$ in the Siegel upper
half-plane $\mathcal H_1=\{\tau\in\IC,\,\Im(\tau)>0\}$.\footnote{The
  complex torus is actually the Jacobian variety of the surface, but at
  genus one both are isomorphic. This property does not hold for
  higher genus curves.} Modding out by the action of the modular group
$SL(2,\IZ)$ restricts $\tau$ to an $SL(2,\IZ)$ fundamental domain. The
one that we use is defined by
${\mathcal F}=\{\tau\in\mathcal{H}_1,\,1<|\tau|,\,-1/2 \leq \Re
\tau<1/2,\, \Im \tau >0\}$, see fig.~\ref{fig:fundtorus}. Also, recall
that 
$${q=\exp({2 i \pi \tau)}}.$$
If we include the three moduli associated to the four punctures at
distinct positions $\zeta_i\in\cT$, $i=1,2,3$ where
$\cT=\{\zeta\in \IC,-1/2<\Re \zeta < 1/2,\,0\leq\Im \zeta < \Im
\tau\}$ and $\zeta_4$ fixed at $\zeta_4=\Im \tau$, we can describe
completely the moduli space $\cM_{1,4}$ over which our string theory
amplitude is integrated
\begin{equation}
\Ast 14 = \int_{\cM_{1,4}}\dmubos~ {\mathcal F}_{1,4}\,.
\label{e:4gravgen1}
\end{equation}
We start the analysis with the four-graviton type II amplitude in $10$
dimensions. Supersymmetry kills the configurations where vertex
operators collide which could create poles. Thus, we will not consider
regions of the moduli space $\Mgn14$ which could give rise to one-loop
diagrams with trees attached to the loop. This will be justified
\textit{a posteriori}. For this amplitude ${\mathcal F}_{1,4}$ is
particularly simple since it is reduced to the Koba-Nielsen factor
times a constant kinematic term
\begin{equation}
  {\mathcal F}_{1,4} = (2 \pi)^8 \mathcal{R}^4  \exp\bigg(\ap\sum_{i<j} k_i\cdot 
  k_j
  \mathcal G(\zeta_i,\bar\zeta_i,\zeta_j,\bar\zeta_j)\bigg)\,,
\label{e:4gravgen1F}
\end{equation}
where $\mathcal{R}^4$ has been defined below eq.~\eqref{e:4gravtree}.
 The integration measure reads
\begin{equation}
\int_{\cM_{1,4}}\dmubos=\int_{\mathcal F} \frac {\d^2\tau }{ (\Im \tau)^5}
\int_{\cT} \prod_{i=1}^3 \d^2 \zeta_i\,.
\label{e:4gravgen1meas}
\end{equation}
The one-loop bosonic propagator reads
\begin{equation}
 \mathcal G(\zeta_i,\bar\zeta_i,\zeta_j,\bar\zeta_j) =-\frac{1}{ 2} 
\log\left|\frac{\theta\charac 1 1 (\zeta_i-\zeta_j|\tau)}
 {\partial_\zeta \theta\charac{1}{1} (0|\tau)} \right|^2 
+\frac{ 2\pi (\Im( \zeta_i-\zeta_j))^2}{\Im \tau}\,,
\label{e:bosproptorus}
\end{equation}
as in \eqref{e:bosprop}. From now on we omit the dependence on the
conjugate variables in $\mathcal G$.
We start the tropicalization procedure, following
sec.~\ref{sec:tropthm}. We look first at the torus alone, and include
punctures later. We want to find a decomposition for ${\mathcal
  F}$. As $q$ is a local coordinate on the moduli space around the
nodal curve at infinity, we would want to use it as in
sec.~\ref{sec:tropicalizing}. We saw in \eqref{e:troplengths} that, in
order to obtain a loop of finite size $T$, we had to set
$|q| = \exp(-2 \pi T/\ap)$. This defines a family of tori parametrized
by their modulus $\tau_\ap$:
\begin{equation}
 \Re\, \tau_\ap = \Re \tau \in[-1/2;1/2[\,,\qquad \Im\, \tau_\ap = T/(2\pi\ap) 
\in [0;+\infty[\,.
\end{equation}
The issue with the previous definition is that for $\Im \tau_\ap<1$,
$\Re \tau_{\ap}$ is not unrestricted in ${\mathcal F}$, but depends on
$\Im \tau_{\ap}$. To build the decomposition, we follow
\cite{Green:1999pv} and introduce a parameter $L>1$ to cut the
fundamental domain into an upper part, the {non-analytic domain}
${\mathcal F}^+(L)$, and a lower part, {the analytic domain}
${\mathcal F}^-(L)$. They are defined by
${\mathcal F}^+(L) = \{\tau\in{\mathcal F},\Im \tau>L\}$ and
$ {\mathcal F}^-(L)=\{\tau\in{\mathcal F},\,\Im \tau \leq L\}$,
respectively.  The decomposition then reads
\begin{equation}
{\mathcal F} = {\mathcal F}^+(L)\sqcup {\mathcal F}^-(L)\,.
\label{e:splitfunddom}
\end{equation}
For any $T\geq 2\pi \ap L$ we now have the standard family of complex
tori in ${\mathcal F}^+(L)$
\begin{equation}
 \Re\, \tau_\ap = \Re \tau \in[-1/2;1/2[\,,\qquad \Im\, \tau_\ap = T/2\pi\ap \in
[L;+\infty[\,.
\label{e:tropscalgen1}
\end{equation}
To complete the decomposition, we have to deal with the
positions of the punctures. Firstly, note that the splitting
\eqref{e:splitfunddom} induces a similar decomposition of $\cM_{1,4}$
into two domains depending on $L$, defined by the position of $\tau$
in ${\mathcal F}$
\begin{equation}
\Mgn14=\Mgn14^+(L) \sqcup \Mgn14^-(L)\,.
\end{equation}
In $\Mgn14^-(L)$, the positions of the punctures can be integrated out
directly. In $\Mgn14^+(L)$ however, it is well known that to take
correctly the $\apt$ limit, one should split the integration domain
spanned the punctures into three regions, one for each inequivalent
ordering of the graph~\cite{Green:1982sw,D'Hoker:1994yr}. Hence
$\Mgn14^+(L)$ is split further into three {disjoint} domains, labeled
by the three permutations inequivalent under reversal symmetry
$\sigma\in\mathfrak S_3/\IZ_2=\{(123),(231),(312)\}$ defined by
\begin{equation}
 \cD_{(ijk)}:= {\mathcal F}(L)^+\, \times\,
\{\zeta_i,\zeta_j,\zeta_k\,|\,0<\Im\zeta_i<\Im\zeta_j<\Im\zeta_k<\Im
\tau\}\,.
\label{e:Dst}
\end{equation}
In total, we have the explicit decomposition
\begin{equation}
 \cM_{1,4}=\bigg(\bigsqcup_{\sigma\in\{(123),(231),(312)\}}
\cD_{\sigma}\bigg)\sqcup \Mgn14^-(L)
\end{equation}
Since the integrand vanishes by supersymmetry in the other regions of
the moduli space, where a tree splits off from the torus for instance,
there is no need to refine the decomposition to take into account
vertex operators colliding to one another.

To determine a tropical form of the integrand, we compute the limit in
the two regions $\Mgn14^\pm(L)$ separately. We define, following
\cite{Green:1999pv},
\begin{equation}
  \begin{aligned}
   & \A14 _{\ap,+} (L)=\sum_{i=(s,t),(t,u),(u,s)} \int_{\cD_i} \dmubos
    {\mathcal F}_{1,4}\,,\\
    &\A14 _{\ap,-} (L)=\int_{\Mgn14^-(L)}\dmutrop\, {\mathcal
      F}_{1,4}\,.
  \end{aligned}
\label{e:lower}
\end{equation}
Of course these partial amplitudes add up to the complete amplitude.

In $\Mgn14^+(L)$, we have the scaling behavior
\eqref{e:tropscalgen1}. As for the punctures, in $\cD_{(ijk)}$ we
define the following families of points:\footnote{This definition is
  equivalent to the one defined in \eqref{e:ptsscal} at tree-level,
  one should just pay attention to the fact that $\zeta_{\ap}$ belongs
  to the complex torus, i.e. the Jacobian. Its inverse image via the
  Abel-Jacobi map, $z_{\alpha'}\simeq\exp (i\zeta_{\alpha'})+O(q)$
  does indeed satisfy \eqref{e:ptsscal}.}
\begin{equation}
 {\zeta_i}_\ap=\Re \zeta_i + i X_i/(2\pi\ap)\,,\quad \Re \zeta_i \in 
[0;2\pi[\,,\quad
0<X_i<X_j<X_k<X_4=T\,.
\label{e:tropchDst}
\end{equation}

Although we already derived in full rigor the field theory limit of
the Green's function at any genus, it is instructive to review this
standard computation at genus one.  The propagator \eqref{e:bosproptorus}
has the following $q$-expansion:
\begin{multline}
    \mathcal G(\zeta_i,\zeta_j) =\frac{ \pi (\Im(
      \zeta_i-\zeta_j))^2}{\Im \tau} -\frac{1}{2} \log\left|\frac{\sin(\pi
      (\zeta_i-\zeta_j))}{ \pi} \right|^2 \\ -2 \sum_{m\geq1}
    \left(\frac{q^m}{ 1-q^m} \frac{\sin^2(m\pi (\zeta_i-\zeta_j))}{ m}
      +h.c.\right)\,,
\label{e:propgen1}
\end{multline}
which, in terms of $\tau_\ap$, ${\zeta_i}_\ap$ and ${\zeta_j}_\ap$,
becomes
\begin{multline}
    \ap \mathcal G({\zeta_i}_\ap,{\zeta_j}_\ap)=\frac 1 {2T}
    (X_i-X_j)^2 -
   \frac {\ap}{2}    
\log\bigg(\big|e^{-(X_i-X_j)/(2\ap)}e^{i\pi\Re(\zeta_{ij})} \\
    -e^{(X_i-X_j)/(2\ap)}
      e^ { -i\pi\Re(\zeta_{ij}) } \big|^2\bigg) +O(\ap)
\label{eq:tropgen1}
\end{multline}
up to $O(q)$ terms and where $\zeta_{ij}$ stands for $\zeta_i-\zeta_j$. At
leading order in $\ap$, the logarithm is equal to the absolute value of
$X_i-X_j$ and one gets
\begin{equation}
 \underset{\apt}\lim(\ap
\mathcal G({\zeta_i}_\ap,{\zeta_j}_\ap))=G^\trop(X_i,X_j)=\frac{1}{2}\left(
- |X_i-X_j|+\frac{(X_i-X_j)^2}{ T}\right)\,.
\label{e:tropgrfct1loop}
\end{equation}
This is the well known worldline propagator on the circle derived in
\cite{Strassler:1992zr} with the exact same normalization. This
expression also coincides with the one for $G^{\trop}$ given in
eq.~\eqref{e:bostropprop}. By plugging that result in
${\mathcal F}_{1,4}$ one obtains
\begin{equation}
{\mathcal F}_{1,4}\to F_{1,4}= (2\pi)^8 \mathcal{R}^4 \exp\left({\sum
k_i\cdot k_j G^{\trop}(X_i,X_j)}\right) + O(\ap)\,,
\label{e:F14trop}
\end{equation}
where nothing depends anymore on the phases $\Re \zeta_i$ or $\Re \tau$. We
can integrate them out and the measure \eqref{e:4gravgen1meas} becomes
\begin{equation}
 \dmubos \to \dmutrop  = 2\pi\ap\frac {\d T }{ T^5}\prod_{i=1}^3 \d X_i
\label{e:tropmeas1loop}
\end{equation}
over the integration domains
 \begin{equation}
D_{(ijk)} = \{T\in[\ap L,+\infty\,[\,\}\times
\{X_i,X_j,X_k
\in [0;\,T[\ |\ 0<X_i<X_j<X_k<T \}\,.
\label{e:Dsttrop}
\end{equation}
For instance in the ordering $1234$, the exponential factor reduces to
$ Q_{1,4} = X_1(X_3-X_2) s + (X_2-X_1)(X_4-X_3) t$; this is the second
Symanzik polynomial of this graph. The first Symanzik polynomial is
simply $T$.

Collecting the various pieces, $\A 14 _{\ap,+}(L)$ is
given by, at leading order;
\begin{eqnarray}\label{e:limitupper}
\A 14 _{+}(L)&=& 
	\sum_{\sigma} \int_{D_\sigma} \dmutrop F_{1,4}\,\\
	&=& \ap (2\pi)^9 \mathcal{R}^4{\bigg(}
		\int^\infty_{2\pi \ap L} \frac{\d T }{ T^2}
		\int_0^{T} \frac{\d X_3 }{ T}
		\int_0^{X_3} \frac{\d X_2 }{ T}
		\int_0^{X_2}\frac {\d X_1 }{ T}\,\,
		e^{Q_{1,4}}\nn\\
		& &\qquad \qquad+{\,2\ \rm other\ orderings}{\bigg)}\,,\nn
\end{eqnarray}
This is the classic result of \cite{Green:1982sw}. Now, we could in
principle drop the restriction $T>2\pi \ap L$ and use dimensional
regularization. However, in order to make the underlying tropical
nature of the limit manifest, the hard UV cut-off $2 \pi \ap L$ should
be kept. Then in 10 dimensions, this integral has a power-behaved
UV-divergence given by
\begin{equation}
  \A 1 4 _{\ap,+} \Big|_{\rm leading~div.} = \ap(2 \pi)^9 \mathcal{R}^4
  \left(\frac{1}{2\pi\ap L}\right)\,,
\label{e:leadingUV}
\end{equation}
as can be seen by a direct computation.
As observed in \cite{Green:1999pv}, the full amplitude $\A 1 4 _{\ap}$
does not depend on $L$, thus any non-vanishing term in
$\A 1 4 _{\ap,+}$ that depends on $L$ in the tropical limit should be
canceled by including contributions from the analytic domain. In
particular, the divergence \eqref{e:leadingUV} should be canceled by a
counter-term coming from $\A 1 4 _{\ap,-}$.

The integrand being analytic in the compact space $\Mgn 1 4 ^-(L)$,
we can take the $\apt$ limit inside the integral: this sets the
exponential factor to~$1$. The integration over the $\zeta_i$'s is now
trivial and the remaining integral can be computed straight away:
\begin{equation}
  \begin{aligned}
    \A 14 _{\ap,-}(L) \to \A 14 _{-}(L) &=(2 \pi)^8 \mathcal{R}^4
    \int_{{\mathcal F}_L} \frac{\d^2\tau}{(\Im \tau)^2}+O(\ap)\\&= 
    (2
    \pi)^9 \mathcal{R}^{4}\left(
      \frac{1}{6}-\frac1{2\pi L}\right)+O(\ap)\,.
  \end{aligned}
\label{e:limitlower}
\end{equation}
Up to the global factor, there are two physically distinct
contributions; $1/6$ and $-1/(2\pi L)$. The first is the so-called
analytic part of the amplitude. After going from the string frame to
the Einstein frame, it is solely expressed in terms of gravitational
coupling constant and is the leading order contribution of higher
order operators in the effective action of supergravity. The second is
the counter-term required to cancel the leading UV divergence
\eqref{e:leadingUV}. From the tropical point of view, this integral
may be thought of as being localized at the singular point $T=0$ of
the tropical moduli space which corresponds to a graph with a vertex
of weight one.

\begin{figure}[t]
\centering
 \begin{center}
 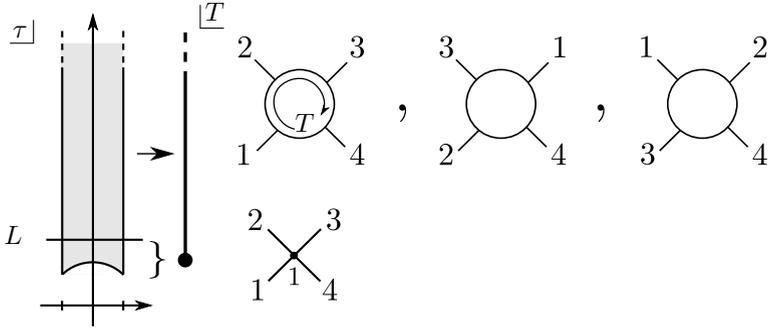
\end{center}
\caption{Summary of the tropicalization of the four-graviton genus one
amplitude in type II string.}
\label{fig:fundtorusL}
\end{figure}

We may now add up \eqref{e:limitupper} and \eqref{e:limitlower} to
obtain the field theory amplitude written as an integral over the full
tropical moduli space $\Mtgn14$. This amplitude is regularized by the
inclusion of a counter-term at $T=0$. This discussion is summarized in
fig~\ref{fig:fundtorusL}.

For general amplitudes, ${\mathcal W}_{1,n}$ acquires a possibly
complicated structure and one often has to perform a Fourier expansion
of $({\mathcal W}_{1,n} \exp({\mathcal Q}_{1,n}))$ in terms of $q$ or
$\sqrt{q}$ as discussed in sec.~\ref{sec:maxim-simpl-maxim} (see
\BKcite and more recently for instance
\cite{Tourkine:2012vx,Tourkine:2012ip} for heterotic string
computations).  At first, these terms may seem $q$- or
$\sqrt{q}$-exponentially suppressed as $\Im \tau\to \infty$.  However,
the worldsheet realization of generic string theory models with
non-maximal supersymmetry is based on altering the spin structure sum
projection: this causes the appearance of ``poles'' in $1/q$ and
$1/\sqrt{q}$. In all consistent models, these poles are automatically
either compensated by higher order terms in the Fourier expansion or
killed by real part integration via identities such as
$\int_{-1/2}^{1/2} q^n\bar q^m \d \Re \tau = 0$ if $n\neq m$.  In the
bosonic string, they are not, which makes the theory inconsistent at
loop level.

Let us make explicit the general form of the decomposition for
$n$-point amplitudes used in the Bern-Kosower rules, or the more
recent works
\cite{BjerrumBohr:2008ji,BjerrumBohr:2008vc,Green:2013bza}. There are
now $(n-1)!/2$ domains $\cD_\sigma$ for
$\sigma \in \mathfrak S_{n-1}/\IZ_2$ defined exactly as in
\eqref{e:Dst} that generate 1PI tropical graphs with orderings
$\sigma$. In this previous analysis we did not have to deal with
regions in the moduli space where points collide to one another
because supersymmetry prevented such configurations to contribute.  In
general though, they have to be included, for physical reasons -- we
know that there are contact terms in generic amplitudes -- and for
mathematical reasons -- the tropical moduli space does have 1PR
graphs.

Hence we refine the previous definition of the domains $\cD_\sigma$
and define new domains $\hat \cD_{\sigma}$ and $\hat\cM^-(L)$ by
cutting out the open discs $|\zeta_i-\zeta_j|<e^{-\ell\ap}$ of the
domains $\cD_\sigma$.\footnote{Note that $\ell$ has to be small
  compared to $L$ so that $\hat \cM^-(L)$ is non-empty.  Typically
  $\ell \ll \sqrt{L/n \pi}$.}  The complementary set of the union of
the previous domains in $\cM^+(L)$ is made of domains of the form
$\hat \cD_\sigma$, where $\sigma \in \mathfrak S_{p-1}/\IZ_2$
indicates the ordering of $p$ points on the future loop while $n-p$
points are grouped into one or more discs of radius $\ell$ centered
around one or more of the first $p$ points.

To finish the description of the decomposition, we have to deal with
these clusters of points. Locally, such a cluster of $m$ points on a
disc of radius $\ell$ looks like a sphere. Thus, as in the tree-level
analysis, $\Mgn1n$ is decomposed into $(2m-3)!!$ domains corresponding
to the $(2m-3)!!$ combinatorially distinct trees. Note the shift
$m\to m+1$ compared to the tree-level case due to the fact that such
trees with $m$ external legs have one additional leg attached to the
loop. 
At this point, one could basically conclude by invoking the
Bern-Kosower rules; this would yield the desired tropical form of the
one-loop amplitude.
Let us then be brief, and describe for simplicity a cluster of two
points, where $\zeta_j$ is treated like before \eqref{e:tropchDst} and
$\zeta_i$ collides to $\zeta_j$;
\begin{equation}\label{e:1PR}
 {\zeta_i}_\ap=\zeta_j+ e^{i\theta}e^{-X/\ap},\quad \theta\in[0;2 \pi[,\quad
X\in[\ap \ell,+\infty[
\end{equation}
where $\zeta_j$ is fixed, $X$ is the tropical length of the tree
connecting legs $i$ and $j$ to the loop as in the tree-level analysis
and $\ell$ is an IR cut-off. In this simple example, there is no outer
region $\cD_0$ and the construction of the decomposition is complete.
Concerning the tropical form of the integrand and the equation
\eqref{prop:sym}, one has to look at
${\mathcal F}_{1,n}={\mathcal W}_{1,n} e^{{\mathcal Q}_{1,n}}$. For
simplicity, we work in a representation of ${\mathcal W}_{1,n}$ where
all double derivatives of the propagator have been integrated out by
parts. Using the general short distance behavior of the propagator on
a generic Riemann surface
\begin{equation}
 \mathcal G(z-w) =-1/2 \log|z-w|^2+ O((z-w)^3)\,,
\end{equation}
one sees that ${\mathcal Q}_{1,n}$ gives a term $-Xk_i\cdot k_j$ while
any term of the form $\mathcal G(\zeta_k,\zeta_i)$ is turned into a $\mathcal 
G(\zeta_k,\zeta_j)$
at leading order in $\ap$:
\begin{equation}
	\sum _{k<l} (k_k\cdot k_l) \mathcal G(kl) = -X (k_i\cdot k_j) 
	+ \sum_{k\neq i,j} k_k\cdot (k_i+k_j) \mathcal G(jk)
	+ \sum_{k<l\atop k,l\neq i,j} (k_k\cdot k_l) \mathcal G(kl)\,,
\end{equation}
up to $O(\alpha')$ terms, with obvious abbreviated notation.
The factor $e^{-X\,k_i\cdot k_j}$ provides a contact term via a pole
in the amplitude if and only if ${\mathcal W}$ contains a factor of
the form $|\partial \mathcal G(ij)|^2\sim e^{2X/\ap}$ exactly as in
the tree-level analysis. Then in ${\mathcal W}$ any
$\zeta_i$-dependent term is replaced by a $\zeta_j$ at the leading
order in $O(\ap)$. This is indeed one of the Bern-Kosower rules.
A similar analysis can be performed in the region $\cM^-(L)$ where we
have to include the contributions of poles. 

In this section, we have recast classic one-loop field theory limits
in the tropical language. This shows a correspondence between the
string theory integration over $\cM_{1,n}$ and its field theory
point-like limit, which can be expressed as an integral over the
tropical moduli space $\Mtgn1n$.
\subsection{Two loops}
\label{sec:two-loop}
Zero- to four-point two-loop amplitudes in RNS type II and heterotic
string have been worked out completely in
\cite{D'Hoker:2001nj,D'Hoker:2001zp,D'Hoker:2001it,
  D'Hoker:2005jb,D'Hoker:2005jc,D'Hoker:2007ui,D'Hoker:2001qp}. The
four-graviton amplitude have also been derived using the pure spinor
formalism \cite{Berkovits:2005df} and shown in \cite{Berkovits:2005ng}
to be equivalent to the RNS computation.

However, the corresponding S-matrix elements in supergravity have not
been extracted from these string theory amplitudes.\footnote{See
  however \cite{D'Hoker:2014gfa}, sec.~3.2, where a degeneration, that
  we call here tropical, of the so-called Kawazumi-Zhang invariant was
  investigated.} In \cite{Green:2008bf}, the four-graviton two-loop
amplitude in maximal supergravity computed in \cite{Bern:1998ug} was
\textit{rewritten} in a worldline form resembling the string theory
integral. In this section, our goal is to prove rigorously that the
tropical limit of the string theory integrand does match this result
by making use of the tropical machinery that we have developed. We
also provide a decomposition of $\Mgn{2}{0}$ such that each region
encompasses the dual graphs corresponding to the primary and sub-
divergences of the amplitude. The study of the integral restricted to
the counter-term domains is left over for future work.

Let us review some facts about genus two Riemann surfaces. At genus
two (and three), there is no Schottky problem, therefore we may
parametrize $\cM_{2}$ in terms of period matrices. As before, the
action of the modular group $Sp(4,\IZ)$ on $\mathcal{H}_{2}$ restricts
it to fundamental domains, of which we pick the representative
$\mathcal{F}_{2}$ defined in~\cite{Gottschling:1959}. This
3-dimensional complex space can be defined in terms of some
inequalities that we describe below. They are similar to these defining
${\mathcal F}$ at genus one. We choose a canonical homology basis
$(a_I,b_J)$ as in figure \ref{fig:canhomolsurf} with normalized
holomorphic one-forms \eqref{e:norm-holdiff}. The period matrix
$\bm \Omega$ is parametrized by three complex moduli $\tau_1,\tau_2$
and $\tau_3$:
\begin{equation}
 \bm \Omega = \begin{pmatrix}
\tau_1+\tau_3&-\tau_3\\
-\tau_3&\tau_2+\tau_3
\end{pmatrix}\,.
\label{e:permatgen2}
\end{equation}
In this parametrization, the inequalities of \cite{Gottschling:1959}
can be rewritten as (see \cite{klkoko});
\begin{itemize}
\item 
Conditions on $\Re \tau_j$ and $\Im \tau_j$:
\begin{equation}    
|\Re \tau_3|\leq \frac{1}{2}, \quad |\Re (\tau_j+\tau_3)|\leq
\frac{1}{2}, \quad     
\Im (\tau_j+\tau_3) \geq\frac{1}{2}\sqrt{3}, \quad j=1,2, \quad \Im
\tau_3 \geq 0    
\label{eq:num1}.
\end{equation}

\item Minkowski ordering: 
\begin{equation}    \Im \tau_{1}\geq \Im \tau_{3}, \quad \Im \tau_{2}\geq \Im 
\tau_{1}
\label{e:mink-ord},
\end{equation}
\item The following set of  19 inequalities:
\begin{equation}    
  |\tau_1+\tau_3|\geq 1,\quad |\tau_2+\tau_3|\geq 1,\quad
  |\tau_1+\tau_2+{\bf \epsilon}|\geq1,    
\label{e:geom-const}
\end{equation}
and
\begin{equation}    
  |\det (\bm \Omega+\bm M)|\geq 1    
\label{e:det-const},
\end{equation}
for all matrices $\bm M$ in the set
\begin{equation}
  \begin{aligned}
  \left\{ ( \begin{smallmatrix} 0 & 0 \\ 0 & 0
      \end{smallmatrix}),  (\begin{smallmatrix} {\bf \epsilon} & 0
        \\ 0 & 0
    \end{smallmatrix}),  (\begin{smallmatrix} 0 & 0 \\ 0 & {\bf
        \epsilon}
    \end{smallmatrix}),  (\begin{smallmatrix} {\bf \epsilon} & 0 \\ 0 & {\bf
        \epsilon}
    \end{smallmatrix}),   (\begin{smallmatrix} {\bf \epsilon} & 0 \\ 0 &
      -{\bf \epsilon}
    \end{smallmatrix}),  (\begin{smallmatrix} 0 & {\bf \epsilon} \\ {\bf
        \epsilon} & 0
    \end{smallmatrix}),  (\begin{smallmatrix} {\bf \epsilon} & {\bf \epsilon} \\
      {\bf \epsilon} & 0
    \end{smallmatrix}),  (\begin{smallmatrix} 0 & {\bf \epsilon} \\ {\bf
        \epsilon} & {\bf \epsilon}
    \end{smallmatrix})  \right\},\, {\bf \epsilon}=\pm 1\,.
    \label{e:set-matrices}
  \end{aligned}
\end{equation}
\end{itemize}

Not considering punctures and ignoring the separating degeneration of
the genus two curve (we will see that it does not contribute to the
field theory limit), we can define a decomposition of $\Mgn{2}{0}$, as
follows. We introduce by hand, in analogy with the genus one
construction, a single parameter $L>1$ and we define three domains
$\cD_i$, $i=a,b,c$ by
\begin{equation}
  \begin{aligned}
    \label{e:domains-genus2}
    \cD_a&= {\mathcal F}_2 \cap \{ \Im \tau_1 \geq L \}\,,\\
    \cD_b&= {\mathcal F}_2 \cap \{ \Im \tau_1 \leq L, \Im \tau_2 \geq L \}\,,\\
    \cD_c&= {\mathcal F}_2 \cap \{ \Im \tau_1 \leq L, \Im \tau_2 \leq L \}\,.
  \end{aligned}
\end{equation}

We checked numerically using a standard numerical minimization routine
that for $L>1$, in the domains $\cD_a$ and $\cD_b$ the determinant
inequalities \eqref{e:det-const} are always satisfied, upon the
constraints eqs.~\eqref{eq:num1}, \eqref{e:geom-const},
\eqref{e:domains-genus2}.  They turn out to be always individually
greater than $L^2$. Of course the same procedure applied in the domain
$\cD_c$ fails for all determinant inequalities, for which the
individual minimums are slightly greater than $0.7$.

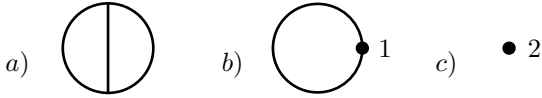
\begin{figure}[t]
\centering{}
$a$)
\parbox{50pt}{ \begin{fmffile}{sunset}
\fmfframe(-4,0)(0,0){ \begin{fmfgraph*}(60,48)
\fmfleft{g1,g2}
\fmfright{g4,g3}
\fmf{phantom}{g1,v1}
\fmf{phantom}{g2,v2}
\fmf{phantom}{g3,v3}
\fmf{phantom}{g4,v4}
\fmf{phantom,tension=0.5}{v1,v2}
\fmf{phantom,tension=0.5}{v2,v3}
\fmf{phantom,tension=0.5}{v3,v4}
\fmf{phantom,tension=0.5}{v4,v1}
\fmf{phantom,left,tension=0.,tag=1}{v1,v3}
\fmf{phantom,left,tension=0.,tag=2}{v3,v1}
\fmf{phantom,left,tension=0.,tag=1}{v2,v4}
\fmf{phantom,left,tension=0.,tag=2}{v4,v2}
\fmfposition
\fmfipath{p[]}
\fmfiset{p1}{vpath1(__v1,__v3)}
\fmfiset{p2}{vpath2(__v3,__v1)}
\fmfi{plain}{subpath (0,length(p1)/2) of p1}
\fmfi{plain}{subpath (length(p1)/2,length(p1)) of p1}
\fmfi{plain}{subpath (0,length(p2)) of p2}
\fmfi{plain}{point 3length(p1)/4 of p1 -- point 3length(p2)/4 of p2}
      \end{fmfgraph*} }
      \end{fmffile} }
\hspace{12pt} $b$)
\parbox{50pt}{ 
		\begin{fmffile}{oneloopct}
		\fmfframe(-8,0)(0,0){ \begin{fmfgraph*}(58,58)
		\fmfleft{g1}
		\fmfright{g2prime,g2,g3,g4,g5}
		\fmf{phantom,tension=1.5}{g1,v1}
		\fmf{phantom,tension=1}{g4,v4}
		\fmf{phantom,tension=1}{g2,v4}
		\fmf{phantom,tension=1}{g3,v4}
		\fmfv{decor.shape=circle,decor.filled=full,
		decor.size=2thick
		}{v4}
		\fmflabel{$1$}{v4}
		\fmf{plain,left,tension=0.2}{v1,v4}
		\fmf{plain,right,tension=0.5}{v1,v4}
      \end{fmfgraph*} }
      \end{fmffile} }
\hspace{12pt}
$c$)
	\parbox{50pt}{ 
		\begin{fmffile}{twoloopct}
		\fmfframe(-4,0)(0,0){ \begin{fmfgraph*}(50,50)
		\fmfleft{g1}
		\fmfright{g4}
		\fmf{phantom,tension=1.5}{g1,v4}
		\fmf{phantom,tension=1}{g4,v4}
		\fmfv{decor.shape=circle,decor.filled=full,
		decor.size=2thick
		,label.dist=-12pt
		}{v4}
		\fmflabel{$2$}{v4}
      \end{fmfgraph*} }
      \end{fmffile} }
\vspace{-12pt}
\caption{From left to right; the three master graphs entering the
  genus two four-graviton amplitude.}
\label{fig:graphs}
\end{figure}

The three domains contain the singularities corresponding to the
graphs of fig.~\ref{fig:graphs}. Therefore, we identify
$\mathcal{D}_a$ as the maximally non-analytic domain and
$\mathcal{D}_c$ as the analytic domain. Since this decomposition is
rather special (as it is defined only in terms of a single parameter
where one could have expected more), it is natural to wonder if the
choice of $L$ is constrained.
Contrary to the one-loop case, the complexity of the definition of the
fundamental domain $\mathcal{F}_2$ does not\textit{ a priori} grant
us that any choice of $L$ would give nice integrals. A good choice for
$L$ would be one that makes the real parts of the $\tau$'s in the
regions $\mathcal{D}_{a}$ and $\mathcal{D}_{b}$ independent from their
imaginary parts, so that they can be integrated out. 
Setting $L$ big enough (of order $10$ for instance) is clearly enough
to ensure that the domain $\mathcal{D}_a$ is of this form, but then it
is not guaranteed that $\mathcal{D}_b$ and $\mathcal{D}_c$ are
suitable for easy integration. In \cite{Pioline:2015nfa} was presented
a more elaborate decomposition based on two parameters, and it would
be interesting to check if it is actually needed for the purpose of
extracting UV divergences and sub-divergences in these amplitudes.

We leave this problem for future investigations, and from now on focus
on the type II four-graviton string amplitude restricted to
$\mathcal{D}_a$, in order to compute the tropical limit of the
integrand.
In ten dimensions it reads
\cite{D'Hoker:2002gw,D'Hoker:2005ht,Berkovits:2005ng,Gomez:2010ad}
\begin{equation}
  \Ast 2 4 (\epsilon_i, k_i)
  =
  \frac{\pi}{64}\left(\frac{\kappa_{10} g_s \ap}{2}\right)^2 \mathcal{R}^4 
  \int_{{\mathcal F}_2} \frac{|\prod_{I\leq J}\d\Omega_{IJ}|^2}
  {(\det\,{\rm Im}\,\bm{\Omega})^5}
  \int_{\Sigma^4}
  |{\mathcal Y}_S|^2
\exp({\mathcal Q}_{2,4})\,.
\label{e:4gravgen2}
\end{equation}
Here, $\int_{\Sigma^4}$ denotes integration of the four punctures over
the surface $\Sigma$. The normalization, in terms of the
10-dimensional gravitational coupling constant $\kappa_{10}$ and the
string coupling constant $g_s$ can be found in \cite{D'Hoker:2013eea}
for instance.
The quantity $\mathcal Y_S$ arises from several contributions in the
RNS computation and from fermionic zero modes in the pure spinor
formalism~\cite{Berkovits:2005df,Berkovits:2005ng}. It
is defined as
\begin{equation}\label{e:Ys}
3  \mathcal Y_S= (k_1-k_2)\cdot (k_3-k_4)\, \Delta(z_1,z_2)\Delta(z_3,z_4)+
(13)(24) + (14)(23)\,,
\end{equation}
with
\begin{equation}
  \Delta(z,w)=\omega_1  (z) \omega_2(w)-\omega_1(w) \omega_2(z)\,,
\label{e:Delta}
\end{equation}
so that $|\Ys|^2$ is a top form on $\Sigma^4$.  Hence we can identify
a measure and an integrand as follows
\begin{subequations}
  \begin{align}
    & \dmubos =\int_{{\mathcal F}_2} \frac{|\prod_{I\leq J}\d\Omega_{IJ}|^2 }{
      ({\rm det}\,{\rm Im}\,\bm \Omega)^5} \int_{\Sigma^4} |{\mathcal{Y}}_S|^2\,\label{e:gen2meas}\,,\\
    &{\mathcal F}_{2,4} = \mathcal{R}^4 \,{\exp}\bigg(\ap \sum_{i<j}k_i\cdot
      k_j\,\mathcal G(z_i,z_j)\bigg),\label{e:gen2F}
  \end{align}
\end{subequations}
where the numerator factor ${\mathcal W}_{2,4}$ is again trivial.

Before starting the computation, we note that it is immediate to see
that the contributions coming from a separating degeneration vanish in
the field theory limit. Indeed, the integrand is missing terms of the
form $\partial G \bar\partial G$ that could produce $1/|z|^{2}$-poles,
required to allow for a massless state exchange. Alternatively, this
can be seen as a consequence of the ``No-triangle'' property of
maximal supergravity,
\cite{BjerrumBohr:2008ji,BjerrumBohr:2008vc}. This justifies why we
did not have to be more precise about this region in defining the
decomposition of $\cM_{2}$.

The degeneration in the domain $\mathcal{D}_{a}$ has already been
studied in details in sec.~\ref{sec:tropicalizing}, around
fig.~\ref{fig:loccoordgen2}. Here we follow a simpler approach: since
we use a parametrization in terms of period matrices, we are allowed
to take the tropical limit directly at this level, instead of at the
level of the curve. Hence, we \textit{define} the tropical
scaling by
\begin{equation}
 \Im \tau_i = -T_i / (2\pi \ap)\,,\quad i=1,2,3\,.
\end{equation}
where, contrary to eq.~\eqref{e:jacscal}, no higher order corrections
enter this equation. Put differently, the $q_{i}$'s, defined by
\begin{equation}
  \label{eq:qidef}
  q_{i}=\exp(2i \pi \tau_{i})
\end{equation}
are particular local coordinates around the boundary divisor which are
only equal to the $t_{i}$'s at leading order ,
${q_i = t_i + O(q_i^{2})}$.  On this point, see \cite[eq
4.6]{Magnea:2013lna} for an explicit relation between the Schottky
representation and the $q_i$ parameters in the case of the genus two
open string worldsheet.

We have thus defined families of curves whose period matrices
tropicalize to $\bm K^{(2)}=\left(\begin{smallmatrix} T_1+T_3&-T_3\\
    -T_3&T_2+T_3 \end{smallmatrix}\right)$.
Furthermore, the boundaries of $\mathcal{D}_{a}$ define worldline
cutoff and ordering given by $\{T_{1}>T_{2}>2\pi \alpha' L,\,T_{3}>0\}$.

Let us now turn to the limit of $\Ys$. The tropical limit of the
holomorphic one-forms \eqref{e:trophd} firstly gives the limit of the
$\Delta$ bilinears;
\begin{equation}
 \Delta(z_i,z_j)\sim \Delta^\trop(ij) = 
\omega_1^\trop(i)\omega_2^\trop(j)-\omega_1^\trop(j)\omega_2^\trop(i)
\end{equation}
up to some factor of $\ap$ that rigorously arises when combining with
the anti-holomorphic part, as in eq.~\eqref{e:riemann-bilin-lim}.
This tropical version of $\Delta$ is defined by
\begin{equation}
\Delta^\trop(ij)=
  \begin{cases}
    ~0  & \text{if }  (i,j) \in B_1~  \text{or } (i,j) \in B_2\\
    ~1 & \text{if }  i\in B_1~  \text{and } j\in B_2\\
    -1 & \text{if }  i\in B_2~  \text{and } j\in B_1
  \end{cases}
\label{e:Deltatrop}
\end{equation}
Then the tropical form of $\Ys$ is immediately obtained:
\begin{equation}
3\Ys\to3Y_S=    (k_1-k_2)\cdot (k_3-k_4)\, \Delta^{\trop}(12)\Delta^{\trop}(34)+
(13)(24) + (14)(23)\,.
\end{equation}
This expression vanishes if three or four punctures lie on the same
edge of the graph, while in all other cases, it is given by a
kinematic invariant as in tab.~\ref{tab:Ys}.
\begin{table}[t]
\begin{tabular}{|c| m{1.5cm} m{1.5cm} m{1.5cm} m{1.5cm}|}
\hline
Graph & \includegraphics[scale=0.8]{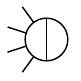} &
\includegraphics[scale=0.8]{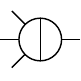}
&
\includegraphics[scale=0.8]{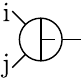} &
\includegraphics[scale=0.8]{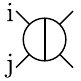}\\
\hline
\multirow{1}{*}{$Y_S$} &\hspace{20pt} $0$ &\hspace{16pt} $0$ &
$(-s_{ij})^{2}$ &$(-s_{ij})^{2}$  $\phantom{\Big(}$\\
\hline
\end{tabular}
\vspace{4pt}
\caption{Numerators for the two-loop four-graviton integrand.}
\vspace{-12pt}
\label{tab:Ys}
\end{table}
Let us mention that $\det \bm K^{(2)} = T_1 T_2+T_2 T_3+T_3 T_1$ does
not depend on the positions of the punctures and is easily seen to be
the usual form of the first Symanzik polynomial of the sunset
graph. This concludes the study of the tropicalization of the
integration measure.

The last thing to be done would be to compute the tropical
representation of the exponential factor \eqref{e:gen2F}.
Fortunately, this was already done at any genus in \eqref{e:expfact},
thanks to theorem~\eqref{e:bostropprop}. Thus we obtain our final
result;
\begin{equation}
  \begin{aligned}
    \A 24 _{\mathrm{non-ana}}(L) =& \mathcal{N} \mathcal{R}^4 \int_{T_1>T_2> 2\pi \ap
      L}^\infty \frac{\d T_1 \d T_2 \d T_3 }{ ({\det} \bm K)^5} \times \\
    &\int_{\Gamma^4} Y_S\, {\exp}\,\bigg(\sum_{i<j}k_i\cdot
    k_j\,G^\trop(Z_i,Z_j)\bigg),
  \end{aligned}
\label{e:4gravgen2trop}
\end{equation}
where $\mathcal{N}$ is a normalization factor and
$\int_{\Gamma^4}$ stands for an integration of the positions of the
four punctures on the graph. This object coincides with the one
derived in \cite[eq. 2.12]{Green:2008bf} from the two-loop field
theory computation of \cite{Bern:1998ug}, thus it is the two-loop
unrenormalized four-graviton amplitude.

To continue the procedure and remove the primary and sub- divergences
(in dimensions when there are any), we should include the
regions $\mathcal{D}_b$ and $\mathcal{D}_c$ described above in
eq.~\eqref{e:domains-genus2}. These computations would illustrate the
systematics of renormalisation in the tropicalization procedure in the
presence of sub-divergences and one should match the field theory
computations of \cite{Green:2008bf,Green:1999pu}.

The computation of the $\apt$ limit of the genus two Heterotic string
amplitude represents a more challenging task, as we said before. It
should be based, as explained in \cite{Tourkine:2012ip}, on a
Fourier expansion of the string integrand in the parameters
$q_i$.

\subsection{A comment at three loops}
\label{sec:comment-three-loops}
An expression was proposed for a sub-sector of the four-graviton
genus-three amplitude using the pure spinor formalism in
\cite{Gomez:2013sla}. Only the terms that contribute to $D^{6}R^{4}$
operator in the low energy limit were computed. Regardless, it would
already be interesting to extract the tropical limit of this partial
amplitude. Comparing the terms obtained from it to the full three-loop
amplitude in supergravity would help to constrain the
form of the missing terms in the string theory computation.
A quick analysis of the tropical limit of this amplitude shows the
following. The integrand of this partial amplitude is a generalization
of the two-loop bilinears $\Delta$ in eq.~\eqref{e:Delta} to
trilinears of the form
$\epsilon_{IJK}\omega^{I}\omega^{J}\omega^{k}$. This kind of terms
always vanish when one $B$-cycle is free of punctures in the tropical
limit, by antisymmetry of $\epsilon_{IJK}$. At the level of the
graphs, this implies, interestingly, that no graph with three or more
particles on the same edge can appear from the 3-loop amplitude, which
is consistent with supersymmetry. However, this also implies that no
``ladder graphs'' can be generated by these terms, since at three
loops the central cycle of ladder graphs is empty. However, such
graphs are definitely present in the three-loop supergravity
amplitude\cite{Bern:2007hh,Bern:2008pv}. Therefore the missing terms
of in the string theory amplitude will have to involve new kind of
objects, different from the $\Delta$'s.

\section{Discussion}
\label{sec:discussion}

The material presented in this paper fits in the active and recent
developments of the domain of string perturbation theory. These are
mostly driven by the introduction of new mathematical structures, for
instance in the automorphic form program
\cite{Green:2010sp,Green:2010kv,Green:2010wi,Green:2011vz,
  Angelantonj:2011br,Angelantonj:2015rxa,Angelantonj:2013eja} or the
analysis of the structure of the supermoduli space
\cite{Witten:2012bh,Witten:2012bg,Witten:2012ga,Witten:2013cia,Witten:2013tpa,
  Witten:2013pra,Donagi:2013dua,Donagi:2014hza,Witten:2015hwa,D'Hoker:2015kwa}
and by certain formal aspects related to genus two and higher string
amplitudes \cite{D'Hoker:2013eea,D'Hoker:2014gfa,Pioline:2015qha,
  Pioline:2014bra,D'Hoker:2013uba,Matone:2012kk,
  Matone:2012rw,Pioline:2015nfa}.  These interactions between physics
and mathematics have yielded significant advances in both domains and
the author hopes that the present work raises some interest in both
communities. \textbf{Note added.} Since this paper appeared on the
arXiv, the author have become aware of the works of Bloch and
collaborators~\cite{Bloch:2015qga,Amini:2015czm}. In these works,
partly inspired by the present paper, the authors describe a
mathematical process very similar to the field theory limit, based on
degenerating mixed Hodge structures. It would be very interesting to
relate precisely the two approaches.

Let us summarize what we achieved in this paper. We formulated the
old-fashioned $\apt$ limit of string theory amplitudes in the context
of tropical geometry: the string theory integral, once split up
according to the domain decomposition \eqref{e:Mgndecomp} provides in
each domain an integral that has the exact same structure as the
expected Feynman integral. By structure, we mean poles inside the
integrand, or equivalently, first and second Symanzik polynomials. The
proof relied on the use of tropical theta functions with
characteristics and on lemma \ref{lem:dist} in particular.
We did not prove that the result of the integration matches
automatically the result obtained from field theory Feynman rules.
This is a separate question, which essentially concerns string field
theory.
We were interested in a practical process that would make use of
pre-computed string theory amplitude and extract the Feynman
numerators in the field theory limit. We reviewed tree and one-loop
processes and performed a two-loop computation. We also commented on
the field theory limit of the three-loop partial amplitude of
\cite{Gomez:2013sla}. This work can be considered as a first step
towards a map between string theory and field theory numerators to all
orders.

Until the recent works of Witten initiated in \cite{Witten:2012bh},
the procedure to compute superstring amplitudes was believed to rely
on the existence of a global holomorphic projection of the supermoduli
space $\sM g n$ onto its bosonic base $\Mgn{g}{n}$
\cite{D'Hoker:2002gw,D'Hoker:1988ta}. It is now known that such a
projection does not exist in general
\cite{Donagi:2013dua,Donagi:2014hza}: for $g\geq5$, $\sM g 0$ is not
holomorphically projected. 
At genus two, the superstring measure (the integrand of the $n=0$
amplitude) was computed in \cite{D'Hoker:2002gw} using an explicit
projection for the even spin structures of $\sM 20$. This result was
obtained by a different method by Witten in \cite{Witten:2013tpa}.
An ansatz at genus three was proposed in \cite{Cacciatori:2008ay},
later extended to genus four in
\cite{Grushevsky:2008qp,Grushevsky:2008zm,Grushevsky:2008zp}.
However, Witten argued~\cite{Witten:2015hwa} that the projection from
the supermoduli space to its bosonic base has a pole in the
bulk of the moduli space (on the hyperelliptic locus), while the ansatz of
\cite{Cacciatori:2008ay} is manifestly holomorphic.

Therefore, the most natural framework for the field theory limit seems
to be a putative super-tropical geometry. The development of such a
theory could eventually allow to treat in full generality first
quantized RNS particles directly on the worldline, and generalize the
seminal work~\cite{Strassler:1992zr}.

Notwithstanding, there are several formulations of string theory that
imply only bosonic integration. For instance the Green Schwarz and the
pure spinor formalisms, but also a few other bosonic realizations of
the superstring~\cite{Ohmori:2013zla}, like that
of~\cite{Berkovits:1993xq}, or topological string amplitudes.
Moreover, the ``vertical integration'' procedure recently introduced by
Sen~\cite{Sen:2014pia,Sen:2015hia} gave a prescription to gauge
fix supergravity on the worldsheet in such a way that the physical
S-matrix elements are independent of this gauge choice. This procedure
is fully generic and allows in principle to perform the integration
over the supermoduli first, using picture changing operators
\cite{Verlinde:1987sd} whose position is integrated using this vertical
integration procedure.

This work was only focussed on the closed string sector. Witten's open
string field theory is based on a particular decomposition of the
moduli space of graphs \cite{Witten:1985cc,Witten:1986qs}, called the
Kontsevich--Penner cell decomposition
\cite{Penner:1987,Kontsevich:1992ti}.\footnote{The author would like
  to thank Edward Witten for pointing out an erroneous use of the
  denomination ``Kontsevich--Penner'' in the first version of this
  draft.} This decomposition describes the moduli space of open string
field theory in terms of proper times \cite{Giddings:1986wp}. It is
different from the one we use here, and it would be interesting to
relate the two. On a related note, in series of works
\cite{DiVecchia:1996kf,DiVecchia:1996uq,Mafra:2012kh,Magnea:2013lna,Magnea:2015fsa},
field theory limits of open string amplitudes have been carefully
studied at one and two loops, using the Schottky parametrization of
Riemann surfaces. The authors of \cite{Magnea:2013lna} also provided
an analysis of the field theory limit in superstring theory based on
super-Schottky parametrization, still in the open string
setting. Inspiration for developing a super-tropical geometry could be
sought in these works.

Another direction for developments how the Feynman $i \epsilon$
prescription fits in the field theory limit. This has been analyzed by
Witten in \cite{Witten:2013pra} where a solution to this question in
string theory was proposed and applied to the description of the field
theory limit of a five-point open bosonic string amplitude restricted
to a specific color ordering $(12345)$.
The moduli space of points on a disk is very similar to $\Mtgn 05$,
except that color ordering selects only one cone through one of the
pentagons, for instance the exterior one in fig.~\ref{fig:M05trop}.
It was shown that the correct string theory integration cycle should
be a complexified version of this cone in order to account for the
$i\epsilon$ prescription (see also \cite{Berera:1992tm}).
Implementing this complexification systematically in the tropical
language would lead to a sort of Lorentzian picture of tropical
graphs. 

Finally, to compute more general tropical limits, it is necessary to
push to higher order the Fourier expansion of the prime
form. In principle, the procedure explained in this paper gives a
prescription for extracting such terms, by choosing the
{appropriate spin structure} -- as in lemma~\ref{lem:dist} --
for each couple of points $(i,j)$ in the factors of
$\partial G(z_i,z_j)$ entering $\mathcal{W}_{g,n}$ to expand the prime
form. The most suited application would be the tropical limit of the
Heterotic string four-graviton two-loop amplitude of
\cite{D'Hoker:2002gw} studied in \cite{Tourkine:2012ip}. Also, the
extraction of the leading and subleading divergences of these two-loop
amplitudes should be performed. An important consistency check of such
a computation is to verify that overlapping and spurious
divergences cancel between the different diagrams. We leave this for
future work.

\subsection*{Acknowledgments}
I would like to acknowledge discussions with several physicists and
mathematicians over the last year. First, let me thank Samuel
Grushevsky for telling me about tropical geometry, as well as for
inspiring discussions. I would also like to thank Ilia Itenberg, Erwan
Brugall\'e, Yan Soibelman and Francis Brown for useful discussions and
remarks. I am also very grateful to Iosif Bena, Guillaume Bossard,
Michael Green, Carlos Mafra, Marco Matone, Stefano Massai, Alexander
Ochirov, Oliver Schlotterer and Kelly Stelle and Pierre Vanhove for
many discussions and remarks.  I am grateful to the organisers of the
Carg\`ese Summer School 2012 where this work was initiated, to these
of the Hamburg Summer School 2013 ``Moduli Spaces in Algebraic
Geometry and Physics'' where parts of this work were achieved and to
the DAMTP and the Niels Bohr International Academy for hospitality at
several stages of this work. Finally I would also like to thank Barak
Kol and Oliver Schlotterer for his very detailed comments on the first
version of the manuscript, Xi Yin for interesting discussions and
Edward Witten for his comments and for pointing out the misuse of the
``Kontsevich-Penner decomposition'' in the first version of this
work. 

Finally, I would like to thank the referee of Annales Henri Poincar\'e
for his detailed and numerous comments on the draft, which greatly
improved its quality.

This research is supported by the Agence Nationale de la Recherche
grant 12-BS05-003-01, from the European Research Council Advanced
grant No. 247252 and from the Centre National de la Recherche
Scientifique grant Projet International de Collaboration Scientifique
6076.

\end{document}